\documentclass[preprint,trackchanges]{aastex62}
\begin{document}


\title{HESS J1943+213: An Extreme Blazar Shining Through The Galactic Plane}



\author{A.~Archer}
\affiliation{Department of Physics, Washington University, St. Louis, MO 63130, USA}

\author{W.~Benbow}
\affiliation{Fred Lawrence Whipple Observatory, Harvard-Smithsonian Center for Astrophysics, Amado, AZ 85645, USA}

\author{R.~Bird}
\affiliation{Department of Physics and Astronomy, University of California, Los Angeles, CA 90095, USA}

\author{R.~Brose}
\affiliation{Institute of Physics and Astronomy, University of Potsdam, 14476 Potsdam-Golm, Germany}
\affiliation{DESY, Platanenallee 6, 15738 Zeuthen, Germany}

\author{M.~Buchovecky}
\affiliation{Department of Physics and Astronomy, University of California, Los Angeles, CA 90095, USA}

\author{V.~Bugaev}
\affiliation{Department of Physics, Washington University, St. Louis, MO 63130, USA}

\author{W.~Cui}
\affiliation{Department of Physics and Astronomy, Purdue University, West Lafayette, IN 47907, USA}
\affiliation{Department of Physics and Center for Astrophysics, Tsinghua University, Beijing 100084, China.}

\author{M.~K.~Daniel}
\affiliation{Fred Lawrence Whipple Observatory, Harvard-Smithsonian Center for Astrophysics, Amado, AZ 85645, USA}

\author{A.~Falcone}
\affiliation{Department of Astronomy and Astrophysics, 525 Davey Lab, Pennsylvania State University, University Park, PA 16802, USA}

\author{Q.~Feng}
\affiliation{Physics Department, McGill University, Montreal, QC H3A 2T8, Canada}

\author{J.~P.~Finley}
\affiliation{Department of Physics and Astronomy, Purdue University, West Lafayette, IN 47907, USA}

\author{A.~Flinders}
\affiliation{Department of Physics and Astronomy, University of Utah, Salt Lake City, UT 84112, USA}

\author{L.~Fortson}
\affiliation{School of Physics and Astronomy, University of Minnesota, Minneapolis, MN 55455, USA}

\author{A.~Furniss}
\affiliation{Department of Physics, California State University - East Bay, Hayward, CA 94542, USA}

\author{G.~H.~Gillanders}
\affiliation{School of Physics, National University of Ireland Galway, University Road, Galway, Ireland}

\author{M.~H\"utten}
\affiliation{DESY, Platanenallee 6, 15738 Zeuthen, Germany}

\author{D.~Hanna}
\affiliation{Physics Department, McGill University, Montreal, QC H3A 2T8, Canada}

\author{O.~Hervet}
\affiliation{Santa Cruz Institute for Particle Physics and Department of Physics, University of California, Santa Cruz, CA 95064, USA}

\author{J.~Holder}
\affiliation{Department of Physics and Astronomy and the Bartol Research Institute, University of Delaware, Newark, DE 19716, USA}

\author{G.~Hughes}
\affiliation{Fred Lawrence Whipple Observatory, Harvard-Smithsonian Center for Astrophysics, Amado, AZ 85645, USA}

\author{T.~B.~Humensky}
\affiliation{Physics Department, Columbia University, New York, NY 10027, USA}

\author{C.~A.~Johnson}
\affiliation{Santa Cruz Institute for Particle Physics and Department of Physics, University of California, Santa Cruz, CA 95064, USA}

\author{P.~Kaaret}
\affiliation{Department of Physics and Astronomy, University of Iowa, Van Allen Hall, Iowa City, IA 52242, USA}

\author{P.~Kar}
\affiliation{Department of Physics and Astronomy, University of Utah, Salt Lake City, UT 84112, USA}

\author{N.~Kelley-Hoskins}
\affiliation{DESY, Platanenallee 6, 15738 Zeuthen, Germany}

\author{D.~Kieda}
\affiliation{Department of Physics and Astronomy, University of Utah, Salt Lake City, UT 84112, USA}

\author{M.~Krause}
\affiliation{DESY, Platanenallee 6, 15738 Zeuthen, Germany}

\author{F.~Krennrich}
\affiliation{Department of Physics and Astronomy, Iowa State University, Ames, IA 50011, USA}

\author{S.~Kumar}
\affiliation{Department of Physics and Astronomy and the Bartol Research Institute, University of Delaware, Newark, DE 19716, USA}

\author{M.~J.~Lang}
\affiliation{School of Physics, National University of Ireland Galway, University Road, Galway, Ireland}

\author{T.~T.Y.~Lin}
\affiliation{Physics Department, McGill University, Montreal, QC H3A 2T8, Canada}

\author{S.~McArthur}
\affiliation{Department of Physics and Astronomy, Purdue University, West Lafayette, IN 47907, USA}

\author{P.~Moriarty}
\affiliation{School of Physics, National University of Ireland Galway, University Road, Galway, Ireland}

\author{R.~Mukherjee}
\affiliation{Department of Physics and Astronomy, Barnard College, Columbia University, NY 10027, USA}

\author{D.~Nieto}
\affiliation{Physics Department, Columbia University, New York, NY 10027, USA}

\author{S.~O'Brien}
\affiliation{School of Physics, University College Dublin, Belfield, Dublin 4, Ireland}

\author{R.~A.~Ong}
\affiliation{Department of Physics and Astronomy, University of California, Los Angeles, CA 90095, USA}

\author{A.~N.~Otte}
\affiliation{School of Physics and Center for Relativistic Astrophysics, Georgia Institute of Technology, 837 State Street NW, Atlanta, GA 30332-0430}

\author{N.~Park}
\affiliation{Enrico Fermi Institute, University of Chicago, Chicago, IL 60637, USA}

\author{A.~Petrashyk}
\affiliation{Physics Department, Columbia University, New York, NY 10027, USA}

\author{M.~Pohl}
\affiliation{Institute of Physics and Astronomy, University of Potsdam, 14476 Potsdam-Golm, Germany}
\affiliation{DESY, Platanenallee 6, 15738 Zeuthen, Germany}

\author{A.~Popkow}
\affiliation{Department of Physics and Astronomy, University of California, Los Angeles, CA 90095, USA}

\author{E.~Pueschel}
\affiliation{DESY, Platanenallee 6, 15738 Zeuthen, Germany}

\author{J.~Quinn}
\affiliation{School of Physics, University College Dublin, Belfield, Dublin 4, Ireland}

\author{K.~Ragan}
\affiliation{Physics Department, McGill University, Montreal, QC H3A 2T8, Canada}

\author{P.~T.~Reynolds}
\affiliation{Department of Physical Sciences, Cork Institute of Technology, Bishopstown, Cork, Ireland}

\author{G.~T.~Richards}
\affiliation{School of Physics and Center for Relativistic Astrophysics, Georgia Institute of Technology, 837 State Street NW, Atlanta, GA 30332-0430}

\author{E.~Roache}
\affiliation{Fred Lawrence Whipple Observatory, Harvard-Smithsonian Center for Astrophysics, Amado, AZ 85645, USA}

\author{C.~Rulten}
\affiliation{School of Physics and Astronomy, University of Minnesota, Minneapolis, MN 55455, USA}

\author{I.~Sadeh}
\affiliation{DESY, Platanenallee 6, 15738 Zeuthen, Germany}

\author{G.~H.~Sembroski}
\affiliation{Department of Physics and Astronomy, Purdue University, West Lafayette, IN 47907, USA}

\author{K.~Shahinyan}
\affiliation{School of Physics and Astronomy, University of Minnesota, Minneapolis, MN 55455, USA}

\author{J.~Tyler}
\affiliation{Physics Department, McGill University, Montreal, QC H3A 2T8, Canada}

\author{S.~P.~Wakely}
\affiliation{Enrico Fermi Institute, University of Chicago, Chicago, IL 60637, USA}

\author{O.~M.~Weiner}
\affiliation{Physics Department, Columbia University, New York, NY 10027, USA}

\author{A.~Weinstein}
\affiliation{Department of Physics and Astronomy, Iowa State University, Ames, IA 50011, USA}

\author{R.~M.~Wells}
\affiliation{Department of Physics and Astronomy, Iowa State University, Ames, IA 50011, USA}

\author{P.~Wilcox}
\affiliation{Department of Physics and Astronomy, University of Iowa, Van Allen Hall, Iowa City, IA 52242, USA}

\author{A.~Wilhelm}
\affiliation{Institute of Physics and Astronomy, University of Potsdam, 14476 Potsdam-Golm, Germany}
\affiliation{DESY, Platanenallee 6, 15738 Zeuthen, Germany}

\author{D.~A.~Williams}
\affiliation{Santa Cruz Institute for Particle Physics and Department of Physics, University of California, Santa Cruz, CA 95064, USA}

\collaboration{The VERITAS Collaboration}

\author{W. F.~Brisken}
\affiliation{Long Baseline Observatory, PO Box O, Socorro, NM 87801, USA}

\author{P.~Pontrelli}
\affiliation{Santa Cruz Institute for Particle Physics and Department of Physics, University of California, Santa Cruz, CA 95064, USA}

\nocollaboration

\correspondingauthor{K.~Shahinyan}
\email{shahin@astro.umn.edu}

\correspondingauthor{O.~Hervet}
\email{ohervet@ucsc.edu}





\begin{abstract}

HESS J1943+213 is a very-high-energy (VHE; $>$100 GeV) $\gamma$-ray source in the direction of the Galactic Plane. Studies exploring the classification of the source are converging towards its identification as an extreme synchrotron BL Lac object. 
Here we present 38 hours of VERITAS observations of HESS J1943+213 taken over two years. The source is detected with $\sim$20 standard deviations significance, showing a remarkably stable flux and spectrum in VHE $\gamma$-rays.
Multi-frequency very-long-baseline array (VLBA) observations of the source confirm the extended, jet-like structure previously found in the 1.6 GHz band with European VLBI Network and detect this component in the 4.6 GHz and the 7.3 GHz bands. The radio spectral indices of the core and the jet and the level of polarization derived from the VLBA observations are in a range typical for blazars. Data from VERITAS, \textit{Fermi}-LAT, \textit{Swift}-XRT, FLWO 48$''$ telescope, and archival infrared and hard X-ray observations are used to construct and model the spectral energy distribution (SED) of the source with a synchrotron-self-Compton model. The well-measured $\gamma$-ray peak of the SED with VERITAS and \textit{Fermi}-LAT provides constraining upper limits on the source redshift. Possible contribution of secondary $\gamma$-rays from ultra-high-energy cosmic ray-initiated electromagnetic cascades to the $\gamma$-ray emission is explored, finding that only a segment of the VHE spectrum can be accommodated with this process. A variability search is performed across X-ray and $\gamma$-ray bands. No statistically significant flux or spectral variability is detected.

\end{abstract}




%
\section{Introduction}
\label{sec:intro}

Blazars are active galactic nuclei (AGN) in which the axis of the relativistic jet is closely aligned with our line of sight \citep{urry95}. The spectral energy distribution (SED) of blazars is characterized by a double-hump structure. In the simpler synchrotron self-Compton (SSC) models, the lower energy hump is attributed to synchrotron emission from relativistic leptons, whereas the higher-energy hump is thought to be from inverse-Compton upscattering of the synchrotron photons on the same population of relativistic leptons~\citep{marscher80,konigl81,reynolds82}. In more complicated scenarios, such as external-Compton models, an external photon field, typically from the accretion disk or the dusty torus around the central black hole is required to explain the higher energy hump~\citep[e.g.,][]{sikora94}. Alternatively, part or all of the $\gamma$-ray emission may be attributed to a hadronic origin, with proton synchrotron radiation and pion production constituting the two primary mechanisms~\citep{mucke01,mucke03}. Variability is a common attribute of blazars, with variations in flux and spectrum detected in every observed frequency band and over a wide range of timescales \citep[see][]{bottcher07}. 

Blazars come in two flavors: BL Lacertae objects (BL Lacs) and Flat Spectrum Radio Quasars (FSRQs), with BL Lacs exhibiting lower power jets and higher Doppler factors and FSRQs possessing high-powered jets and showing high Compton dominance~\citep{stickel91,stickel93}. Based on the location of the synchrotron peak, BL Lacs are classified into low, intermediate, and high synchrotron peak BL Lacs (LBLs, IBLs, and HBLs respectively)~\citep{padovani95}. HBLs are the most commonly detected blazars in VHE $\gamma$-rays, comprising 47 of the 64 VHE-detected blazars\footnote{tevcat2.uchicago.edu}. A subclass of HBLs has been proposed, known as extreme HBLs (EHBLs), identified by synchrotron emission peaks at energies above 1~keV \citep{costamante01}. 

Within the context of the blazar sequence \citep[e.g.,][]{fossati98,ghisellini08,meyer11,giommi12} -- where the blazar jet luminosity is inversely related to the Doppler factor -- EHBLs would be the least luminous and would have the highest Doppler boosting factors, making them one of the most efficient and extreme accelerators in the Universe. However, with only a handful of blazars belonging to the EHBL subclass (including 1ES 0229+200, 1ES 0347-121, RGB J0710+591, 1ES 1101-232), there is as yet no conclusive physical explanation for them. 

EHBLs constitute a challenge for leptonic emission models that tend to only accommodate the observed spectral energy distributions of these objects with unusually hard particle populations~\citep[e.g.,][]{tavecchio10}. Moreover, unlike other blazars, they do not appear to exhibit rapid variability, despite predictions of large flux variations on short timescales by leptonic models. The higher synchrotron peak frequency could potentially explain this as an observational effect by shifting the more variable emission produced by higher energy particles into the hard X-ray band. The less energetic particles producing steadier emission would then be responsible for the emission in the commonly observed infrared to soft X-ray bands.

The lack of rapid flux variability and the hard VHE spectra make EHBLs attractive candidates for hadronic emission models. Their observed properties can be explained by synchrotron emission from relativistic protons within the jet and by proton-initiated electromagnetic cascades~\citep{cerruti2015}. As such, the more distant EHBLs are also ideal candidates for a proposed $\gamma$-ray emission mechanism in which at least a component of the observed VHE emission originates from ultra-high-energy cosmic rays (UHECRs) that propagate an appreciable fraction of the distance between the blazar and Earth before producing electromagnetic cascades along the line of sight ~\citep[see][]{ferrigno04, bonnoli15, essey10}. If either mechanism is confirmed, EHBLs would become one of the most likely sources for the acceleration sites of UHECRs, directly addressing one of the oldest questions in high-energy astrophysics. 

\subsection{HESS J1943+213: An Extreme HBL}

\objectname{HESS J1943+213} is a VHE $\gamma$-ray point source discovered during the H.E.S.S. Galactic Plane scan~\citep{hess11}. Since the discovery publication, the identity of HESS J1943+213 has been a topic of debate, with most of the observations suggesting the source is a blazar, but with alternative possibilities including a pulsar wind nebula (PWN) and a $\gamma$-ray binary.

Assuming the source is a $\gamma$-ray binary, \citet{hess11} used the lack of detection of a massive (O- or Be-type) companion star to estimate a distance limit of greater than $\sim$25 kpc. This distance places the binary well beyond the extent of the Galactic disk and implies an X-ray luminosity 100--1000 times higher than luminosities of known $\gamma$-ray binaries. Hence, \citet{hess11} disfavor the $\gamma$-ray binary scenario. 

The point-like appearance in X-rays and the soft VHE spectrum, with a power-law index of $\Gamma$ = 3.1 $\pm$ 0.3, motivated ~\citet{hess11} to argue against the PWN scenario. However, 1.6-GHz observations of the HESS J1943+213 counterpart with the European VLBI Network (EVN) detected an extended source, with FWHM angular size of 15.7~mas (with 3.5~mas expected size for a point source)~\citep{gabanyi13}. Based on this measurement, the brightness temperature of the counterpart was estimated to be 7.7$\times$10$^{7}$ K and was used to argue against the blazar scenario, as the expected brightness temperature of VHE-detected HBLs is in the 10$^{9}$--10$^{10}$ K range. In addition, \citet{gabanyi13} employed a 1$'$ feature observed in the 1.4-GHz very-large array (VLA) C-array configuration image to support the PWN hypothesis, with the assertion that the angular size of the feature is consistent with a Crab-like PWN placed at a distance of 17~kpc. However, a pulsar search with the Arecibo telescope resulted in a non-detection and a claim of no pulsar at the HESS J1943+213 location at $\sim$70$\%$ confidence~\citep{straal16}.

As reported by \citet{hess11}, all observations were found to be consistent with the blazar scenario, however, including the point-like nature in both X-rays and VHE, the soft VHE spectral index, and an (unpublished) featureless IR spectrum. In addition,~\citet{tanaka14} argued in favor of an EHBL by constructing a spectral energy distribution and by drawing comparisons to a known EHBL, 1ES 0347-121. The case for the extreme blazar has been bolstered further with~\citet{peter14} observing the near-infrared (K-band) counterpart of HESS J1943+213 and claiming a detection of an elliptical host galaxy.

Recently, ~\citet{straal16} obtained VLBI observations in the 1.5 GHz and 5 GHz bands using the e-Multiple Element Remotely Linked Interferometer Network (e-MERLIN), showing that the source exhibits a flat spectrum between the two bands and claiming a detection of flux density variability in the 1.5 GHz band when compared with the EVN observations of the source. 

A strong argument for the blazar case was made with a reanalysis of the initial EVN dataset and addition of new and higher resolution observations in 2014~\citep{akiyama16}. Based on both EVN observations, the brightness temperature of the core is estimated to be well within the blazar range with T${_B}$ $>$  1.8$\times$10$^{9}$~K  and  7.7$\times$10$^{9}$~K for 2011 and 2014 observations respectively. The claim for flux density variability was also made more robust through a consistent analysis of EVN data from the two epochs. In addition, the 2014 EVN observations revealed extended jet-like structure with brightness temperatures of the individual substructures of the extended emission typical of AGN jets.

The arguments presented above strongly suggest that HESS J1943+213 is a BL Lac object behind the Galactic Plane. With a synchrotron peak located at $\sim$10 keV and with no apparent cutoff, HESS J1943+213 is classified as an extreme synchrotron BL Lac object or an EHBL. In addition to the location of the synchrotron peak, HESS J1943+213 displays other attributes of EHBLs, including a very large X-ray to radio flux ratio, weak emission in the GeV band and a lack of strong flux variability relative to other blazars.

There are only indirect limits on the distance of HESS J1943+213 measured by \citet{peter14}. Lower limits on the redshift come from the assumed size for the host galaxy and measurement of its extension in near IR, while upper limits are derived by extrapolating the \textit{Fermi}-LAT spectrum into the VHE regime and assuming that the deviations from the extrapolated spectrum are entirely due to absorption by the extragalactic background light (EBL). The redshift bounds found by \citet{peter14} are 0.03 $<$ z $<$ 0.45. 

In what follows, we detail results from observations of HESS J1943+213 with VERITAS, \textit{Fermi}-LAT, \textit{Swift}-XRT,  and VLBA and further characterize the properties of the source as an EHBL. Section~\ref{sec:data} presents new observations and results collected with VERITAS and VLBA, in addition to analyses of 8-years of \textit{Fermi}-LAT data, recent \textit{Swift}-XRT observations, and long-term optical observations with the Fred Lawrence Whipple Observatory (FLWO) 48$''$ telescope. The results from the analysis of HESS J1943+213 multi-wavelength data are interpreted and discussed in Section~\ref{sec:interp}, including a derivation of improved and more robust limits on the source redshift (\ref{sec:redshift}) based on the $\gamma$-ray spectra from \textit{Fermi}-LAT and VERITAS. We perform a search for variability in X-ray and $\gamma$-ray observations of HESS J1943+213 (\ref{sec:var}), construct and model the broadband spectral energy distribution of the source (\ref{sec:sed}), and explore UHECR line-of-sight $\gamma$-ray production as an alternative emission mechanism (\ref{sec:crpropa}). We conclude in Section~\ref{sec:conclusions}.

\section{Multi-wavelength Observations of HESS J1943+213 and Data Analysis}
\label{sec:data}

\subsection{Strong Detection and Characterization of the Source with VERITAS}
\label{sec:veritas}

The Very Energetic Radiation Telescope Array System (VERITAS) is an imaging atmospheric Cherenkov 
telescope array located at the Fred Lawrence Whipple Observatory in southern Arizona (31$^{\circ}$ 40$'$ N, 110$^{\circ}$ 57$'$ W,  1.3~km a.s.l.). VERITAS is composed of four 12-m telescopes, each equipped with a 499 photo-multiplier tube camera providing a 3.5$^{\circ}$ field of view \citep{holder06}. The array can reliably reconstruct $\gamma$-rays with energies between 85~GeV and 30~TeV, and has an angular resolution at 68$\%$ containment of $<$ 0.1 degrees for a 1~TeV photon. The energy resolution is 17$\%$ at 1~TeV, with a 10$^{5}$ m$^{2}$ peak effective area \citep{park15}.

VERITAS observations of HESS J1943+213 took place over $\sim$2.5 years and are broken up into two periods for spectral analysis: 
(I) 2014 May 27 (MJD 56804) -- 2014 July 02 (MJD 56840), (II) 2015 April 20 (MJD 57132) -- 2015 November 09 (MJD 57335). The total exposure time of these observations
is 37.2~hours, amounting to a weather-cleaned live time of 30.9~hours. Observations from Period I focused on deep exposures of the source and constitute 24.2 hours of weather-cleaned data, while the remaining 6.7 hours in Period II aimed at probing  the source for variability. The source elevation during the VERITAS observations was within the 
63$^{\circ}$ -- 80$^{\circ}$ range, with the common low-energy threshold for this analysis determined to be 180~GeV.
 
The analysis of the VERITAS data is performed and cross-evaluated for consistency with the two independent, standard VERITAS analysis packages~\citep{vegas,daniel08}. The images of Cherenkov light from particle showers are parameterized with the classical Hillas approach~\citep{hillas85}. Standard cuts optimized for average source strength ($\sim$5\% Crab Nebula flux) and spectral index ($\Gamma \sim 2.7$) are used for separating $\gamma$-ray and cosmic ray events~\citep[See][for details]{Acciari08}. The background for $\gamma$-ray-like events is measured using the reflected regions method~\citep{fomin94}. The source significance is calculated using the generalized version of equation 17 from~\citet{LiMa} derived by~\citet{klepser12}.

\floattable
\begin{deluxetable}{ccccccccc}
\tabletypesize{\scriptsize}
\tablecaption{\small Summary of VERITAS observations.\label{tab:ver}}
\tablewidth{0pt}
\tablehead{
\colhead{Period} & \colhead{Exposure} & \colhead{On Counts} & \colhead{Off Counts} & \colhead{$\alpha$$^{\text{a}}$} & \colhead{$\sigma$} & \colhead{Flux ($>$180~GeV)} & \colhead{$\Gamma$} & \colhead{$\chi$$^{2}$/NDF} \\
& \colhead{[hours]} & & & & & \colhead{[cm$^{-2}$ s$^{-1}$]} & & 
}
\startdata	
I & 24.2 & 713  & 7684 & 0.043 & 17.9 & (8.61$\pm$0.78)$\times$10$^{-12}$ & 2.76$\pm$0.12 & 3.5/6 \\
II & 6.7 & 164 & 2087 & 0.042 & 7.2 & (8.55$\pm$1.88)$\times$10$^{-12}$ & 3.12$\pm$0.38 & 4.4/4 \\
Combined & 30.9 & 877 & 9771 & 0.042 & 19.3 & (8.61$\pm$0.67)$\times$10$^{-12}$ & 2.81$\pm$0.12 & 3.8/5 \\
\enddata
\tablenotetext{\text{a}}{Off region-source normalization.}
\end{deluxetable}

A source centered at $\alpha$ = 19$^{h}$43$^{m}$59$^{s}$ $\pm$ 1$^{s}$$_{\text{(stat)}}$ $\pm$ 2$^{s}$$_{\text{(sys)}}$ and 
$\delta$ = 21$^{\circ}$19$'$05$''$ $\pm$ 11$''_{\text{(stat)}}$ $\pm$ 25$''_{\text{(sys)}}$ is detected with an excess of 19.3 standard deviations ($\sigma$), 
consistent with the catalog position of HESS J1943+213. The VERITAS detection is fit by a 2-dimensional Gaussian function representing the VERITAS point spread function. The fit $\chi$$^2$/NDF is 2069/1931, which corresponds to a fit probability of 1.5\%. The best-fit Gaussian width is 0.05 degrees, smaller than the angular resolution of VERITAS. Thus, there is no evidence for source extension. The VERITAS source name is \objectname{VER J1943+213}.

\begin{figure}[ht]
\plotone{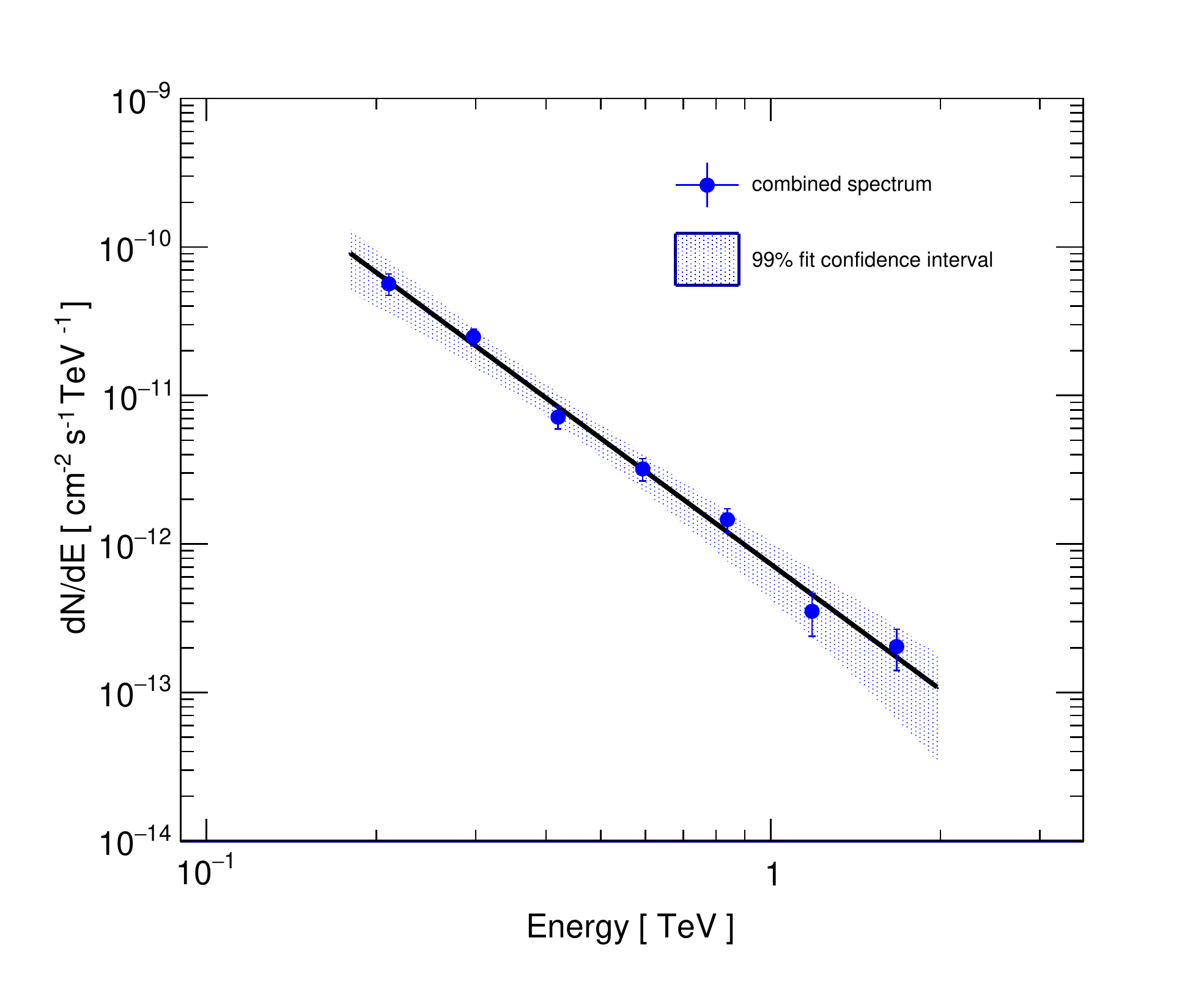}
\caption{Time-averaged VERITAS spectrum of HESS J1943+213, combining data from 2014 and 2015 observations. The band shows the 99\% confidence interval of a power-law fit to the spectrum. \label{fig:veritas_spectra}}
\end{figure}

\begin{figure}[ht]
\plotone{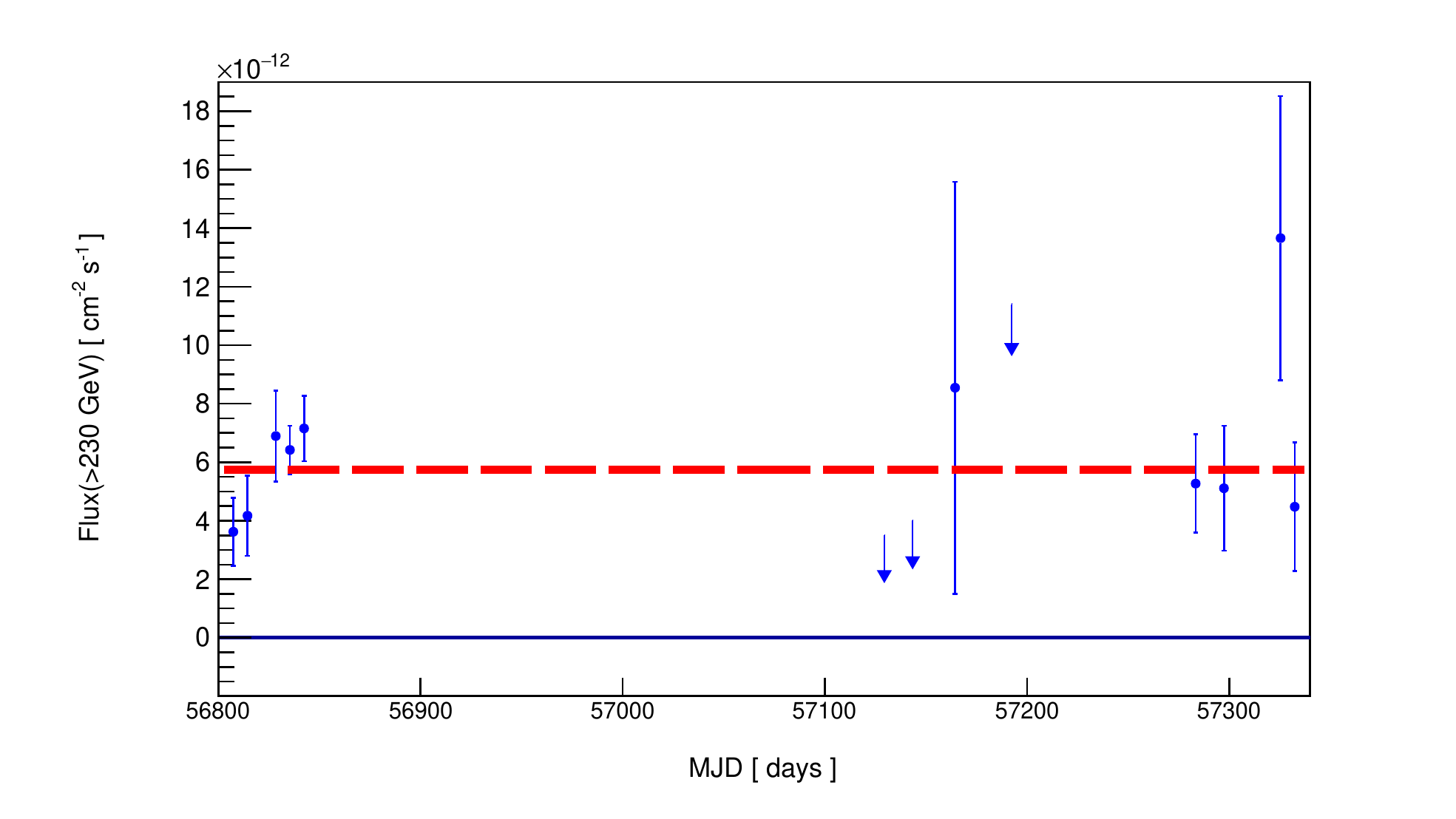}
\caption{VERITAS light curve of HESS J1943+213 above 230~GeV. The red dashed line is a fit of a constant to the data. 68\% upper limits are derived for time bins in which the source flux is consistent with zero.\label{fig:veritas_lightcurve}}
\end{figure}

The VERITAS differential energy spectra of HESS J1943+213 are constructed separately for each period
specified above to look for variations in the source spectrum. The spectra from
the two periods agree with each other within the statistical uncertainties, indicating no significant detection of spectral
variability and justifying the use of a combined, time-averaged spectrum from the entire dataset for the SED modeling detailed in 
Section~\ref{sec:sed}. The combined VERITAS spectrum of HESS J1943+213 is presented in Figure~\ref{fig:veritas_spectra}. The spectrum is fit well by a power-law function, $\frac{dN}{dE}$ = N$_{0}$\big($\frac{E}{E_{0}}$\big)$^{-\Gamma}$, with a spectral index, $\Gamma$ = 2.81$\pm$0.12$_{\text{(stat)}}$$^{+0.14}_{-0.34}$$_{\text{(sys)}}$ in the 180~GeV--2~TeV energy range. \citet{hess11} reported a spectral index of 3.1$\pm$0.3$_{\text{(stat)}}$$\pm$0.2$_{\text{(sys)}}$ for HESS J1943+213 above 470 GeV. In the same energy range, the VERITAS best-fit spectral index is 2.85$\pm$0.32$_{\text{(stat)}}$$^{+0.14}_{-0.34}$$_{\text{(sys)}}$, consistent with the H.E.S.S. result.

There is no evidence for flux variability between Periods I and II and on
weekly timescales as illustrated by the VERITAS light curve of HESS J1943+213 in Figure~\ref{fig:veritas_lightcurve}.  The average flux of the source is (5.57$\pm$0.46$_{\text{(stat)}}$$_{-1.27}^{+0.72}$$_{\text{(sys)}}$)$\times$10$^{-12}$ cm$^{-2}$ s$^{-1}$ above 230 GeV. A constant line is fit to the full weekly-binned light curve with $\chi^2$/NDF = 15.6/12, corresponding to a p-value of 0.21 for a constant flux. The higher cadence Period I observations were investigated for variability separately on both daily and weekly timescales. No significant variability was found in that dataset, with $\chi^2$/NDF = 31.34/20 for a constant fit to a daily light curve, corresponding to a p-value of 0.05 for a constant flux~\citep{shahinyan15}.

H.E.S.S. reported a source flux of (1.3$\pm$0.2$_{\text{(stat)}}$$\pm$0.3$_{\text{(sys)}}$)$\times$10$^{-12}$ cm$^{-2}$ s$^{-1}$
 above 470~GeV \citep{hess11} from observations taken between 2005 and 2008. This is consistent with the VERITAS flux above 470~GeV of (1.47$\pm$0.16$_{\text{(stat)}}$$_{-1.27}^{+0.72}$$_{\text{(sys)}}$)$\times$10$^{-12}$ cm$^{-2}$ s$^{-1}$. Thus, in addition to the remarkable stability of the source flux over two years of VERITAS observations, there is also good agreement between fluxes from observations more than six years apart from two different VHE $\gamma$-ray instruments.

\clearpage
\subsection{Improved Detection and Spectral Analysis with an 8-year \textit{Fermi}-LAT Dataset}
\label{sec:lat}

The Large Area Telescope (LAT) on board the \textit{Fermi} Gamma-ray Space Telescope (\textit{Fermi}-LAT) is a pair-conversion $\gamma$-ray instrument sensitive to high energy (HE) $\gamma$-rays with energies between 20~MeV and 300~GeV. The LAT has a field of view that covers $\sim$20$\%$ of the sky at any given time in survey mode, providing 30 minutes of live time on each point in the sky every two orbits ($\sim$3 hours).

A source associated with HESS J1943+213 is included in the Second Catalog of Hard \textit{Fermi}-LAT Sources (2FHL) \citep{2fhl} within the 50~GeV -- 2~TeV energy range, with a TS = 39.6 and a power-law spectral index of 2.73 $\pm$ 0.66. In addition, the source is included in the preliminary Third Catalog of Hard \textit{Fermi}-LAT Sources (3FHL)~\citep{3fhl} in the 10~GeV -- 2~TeV range, with a TS = 127.7 and a spectral index of 1.45 $\pm$ 0.29. HESS J1943+213 has been previously detected at 5.1 $\sigma$ significance through an analysis of 5 years of \textit{Fermi}-LAT data in the 1--100~GeV energy range, producing a spectrum well fit by a power-law with normalization of (3.0 $\pm$ 0.8$_{\text{(stat)}}$ $\pm$ 0.6$_{\text{(sys)}}$) $\times$ 10$^{-15}$ cm$^{2}$ s$^{-1}$ MeV$^{-1}$ at 15.1~GeV and spectral index, $\Gamma$ = 1.59 $\pm$ 0.19$_{\text{(stat)}}$ $\pm$ 0.13$_{\text{(sys)}}$ \citep{peter14}.

In this work, the Fermi Science Tools\footnote{https://fermi.gsfc.nasa.gov/ssc/data/analysis} version \textit{v10r0p5} and the P8R2\_SOURCE\_V6 instrument response functions are used for analyzing \textit{Fermi}-LAT observations, with the assistance of Fermipy~\citep{wood17} - a python package with a high-level interface for Fermi-LAT analysis..
Eight years of PASS8 \textit{Fermi}-LAT data are selected for the analysis between 2008 August 04 and 2016 August 04. The region of interest (RoI) is defined within a 10$^\circ$ radius of the catalog position of HESS J1943+213 ($\alpha$: 19$^{h}$43$^{m}$55$^{s}$, $\delta$: 21$^{\circ}$18$'$8$''$). SOURCE class events with energies in the 3 -- 300~GeV range are selected. The 3~GeV lower bound on the energy is chosen to decrease the contribution from the Galactic diffuse background. In addition, only events with zenith angles $<$ 100$^{\circ}$ and rocking angles $<$ 52$^{\circ}$ are included to avoid contamination from the Earth limb. 

A model for the RoI is constructed by including all \textit{Fermi}-LAT Third Source Catalog (3FGL) \citep{3fgl} sources within 15$^\circ$ radius of the source position and models for emission from the Galactic diffuse (\textit{gll\_iem\_v06.fits}) and the isotropic (\textit{iso\_P8R2\_SOURCE\_V6\_v06.txt}) backgrounds. In addition, a point source with a power-law spectrum is added at the catalog position of HESS J1943+213. A binned likelihood analysis is performed to find the optimal model for the RoI and extract the best-fit source parameters. The parameters of weak sources with test statistic (TS) $<$ 16 and all sources located more than 7$^{\circ}$ away from the center of the RoI are frozen during this procedure.

A source at the position of HESS J1943+213 is detected with a TS value of 147.5 corresponding to a significance of $\sim$12 $\sigma$. The source is modeled as a power-law function with an index of 1.67$\pm$0.11$_{\text{(stat)}}$ and a flux of (2.71$\pm$0.43$_{\text{(stat)}}$)$\times$10$^{-10}$ cm$^{-2}$ s$^{-1}$ above 3~GeV.
Spectral points are generated by repeating the \textit{Fermi}-LAT analysis with events selected within the energy range of each spectral bin. In addition, different spectral shapes are explored for the HESS J1943+213 detection showing no statistically significant preference for a curved model over a power law.

\subsection{\textit{Swift}-XRT Observations Contemporanous with VERITAS}
\label{sec:xrt}

X-ray observations of HESS J1943+213 were obtained with the \textit{Swift}-XRT instrument on 2014 June 17, 2014 June 19, and 2014 June 21, contemporaneous with VERITAS observations. The XRT data analyzed here were collected in the \textit{photon counting} mode, amounting to a total of 48.2 minutes of exposure time. 

The XRT data analysis is performed with the standard XRTDAS v3.0.0 tools included in the HEASoft package Version 6.15.1, while Xspec~\citep{arnaud96} v12.8.1g is used for the spectral analysis.

\floattable
\begin{deluxetable}{cccccc}
\tabletypesize{\scriptsize}
\tablecaption{\small Summary of \textit{Swift}--XRT observations.\label{tab:xrt}}
\tablewidth{0pt}
\tablehead{
\colhead{Date} & \colhead{Observation ID} & \colhead{Exposure} & \colhead{Log10 [Flux] (2--10 keV)} & \colhead{Index} & \colhead{$\chi$$^{2}$/NDF} \\
& & \colhead{[seconds]} & \colhead{Log10 [erg cm$^{-2}$ s$^{-1}$]} & &
}
\startdata	
2014 June 17 & 00033319001 & 967 &-10.75$\pm$0.05 & 2.16$\pm$0.18 & 4.23/9 \\
2014 June 19 & 00033319002 & 769  &-10.66$\pm$0.05 & 1.92$\pm$0.16 & 3.29/8 \\
2014 June 21 & 00033319003 & 1156 & -10.53$\pm$0.04 & 1.77$\pm$0.13 & 12.15/17 \\
\enddata
\end{deluxetable}

XRT spectra are constructed by unfolding the counts spectra with instrument response functions 
included in CALDB 1.0.2 and by assuming an absorbed power-law functional form for the intrinsic spectrum: $\frac{dN}{dE}$ = N$_{0}$\big($\frac{E}{E_{0}}$\big)$^{-\Gamma}$$e^{-\text{N}_{\text{H}}\sigma(E)}$, where $\sigma$(E) is the photo-electric cross-section and N$_{\text{H}}$ is the HI column density. Spectra were first fit using absorbed power-law functions with the N$_{\text{H}}$ parameter left free to search for excess over the Galactic N$_{\text{H}}$ value obtained from the Leiden/Argentine/Bonn (LAB) survey, N$_{\text{H}}$ $\sim$ 0.82$\times$10$^{22}$ cm$^{-2}$ \citep{labsurvey}. The fitted value of N$_{\text{H}}$ is consistent with the value from the LAB survey for the observation on 2014 June 17 (N$_{\text{H}}$ = (0.87$\pm$0.48)$\times$10$^{22}$ cm$^{-2}$) and slightly exceeds the LAB survey value for observations on 2014 June 19 (N$_{\text{H}}$ = (1.48$\pm$0.54)$\times$10$^{22}$ cm$^{-2}$) and 2014 June 21 (N$_{\text{H}}$ = (1.21$\pm$0.35)$\times$10$^{22}$ cm$^{-2}$). The high-precision \textit{Suzaku}-XIS and HXD/PIN spectrum from \citet{tanaka14} measured N$_{\text{H}}$ = (1.38 $\pm$ 0.03)$\times$10$^{22}$ cm$^{-2}$, also in excess of the LAB survey value.

\begin{figure}[ht]
\plotone{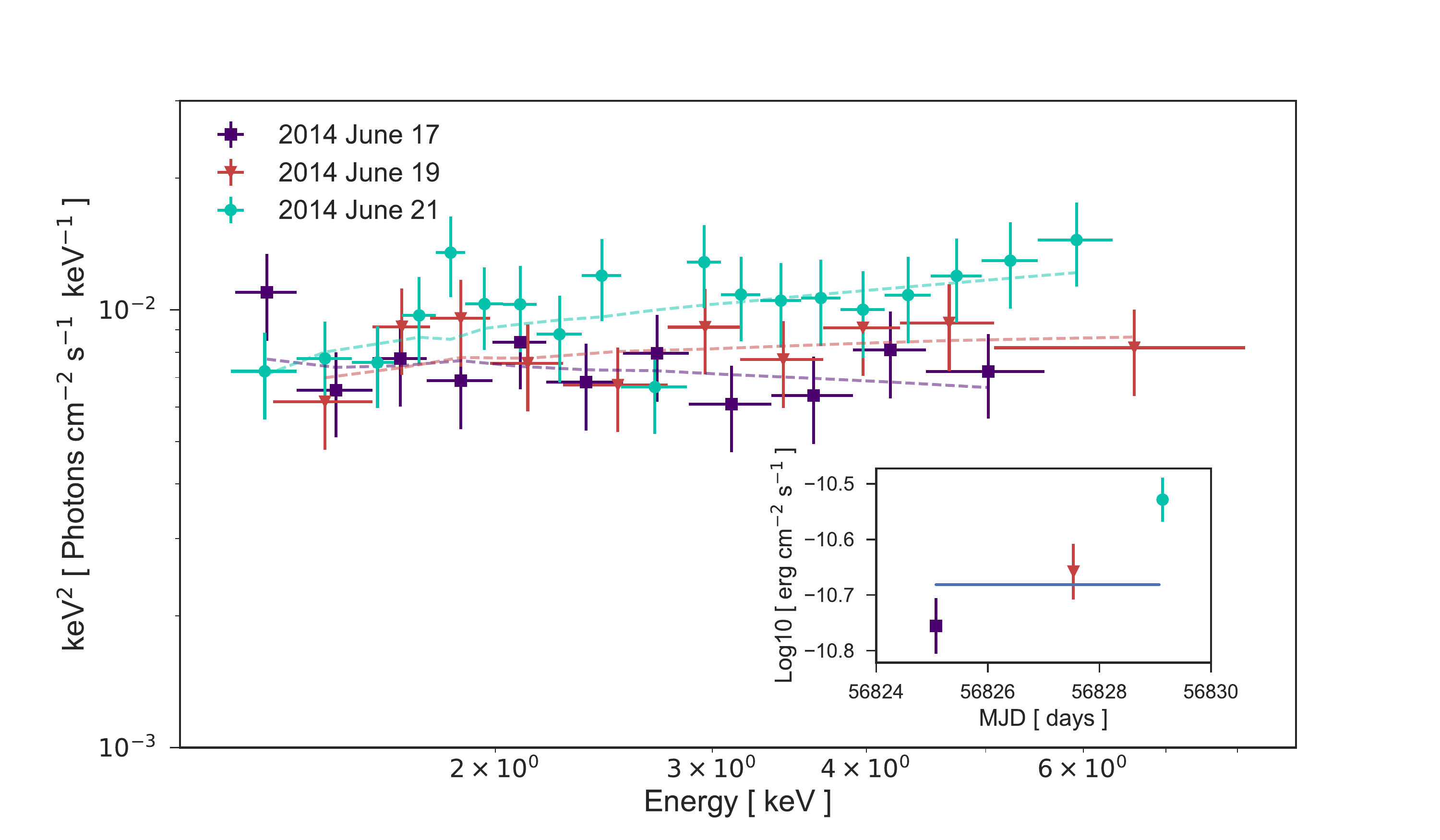}
\caption{Spectra of HESS J1943+213 with \textit{Swift}-XRT for the three observations taken in 2014. The dashed lines show the absorbed power-law models used for unfolding the spectra. The inset figure shows the 2--10~keV fluxes for the three observations.}
\label{fig:xrt_spectra}
\end{figure}

The spectrum from each observation is shown in Figure \ref{fig:xrt_spectra}. Spectra are again constructed assuming absorbed power-law functional forms; however, the N$_{\text{H}}$ parameter is kept fixed to the value from \citet{tanaka14}. The results from the spectral fits are included in Table~\ref{tab:xrt}. The uncertainties represent the 68\% confidence intervals for the respective quantities. No significant spectral variability is detected between observations. Moreover, the results are comparable to measurements from 2006 October 10 observations of HESS J1943+213 with \emph{Swift}-XRT, which reported a spectral index of 2.04 $\pm$ 0.12 and a flux of (1.83 $\pm$ 0.04) $\times$ 10$^{-11}$ erg cm$^{-2}$ s$^{-1}$ \citep{malizia07, landi09}.

\subsection{Long-Term Optical Observations with FLWO 48$''$}

\begin{figure}[ht]
\plotone{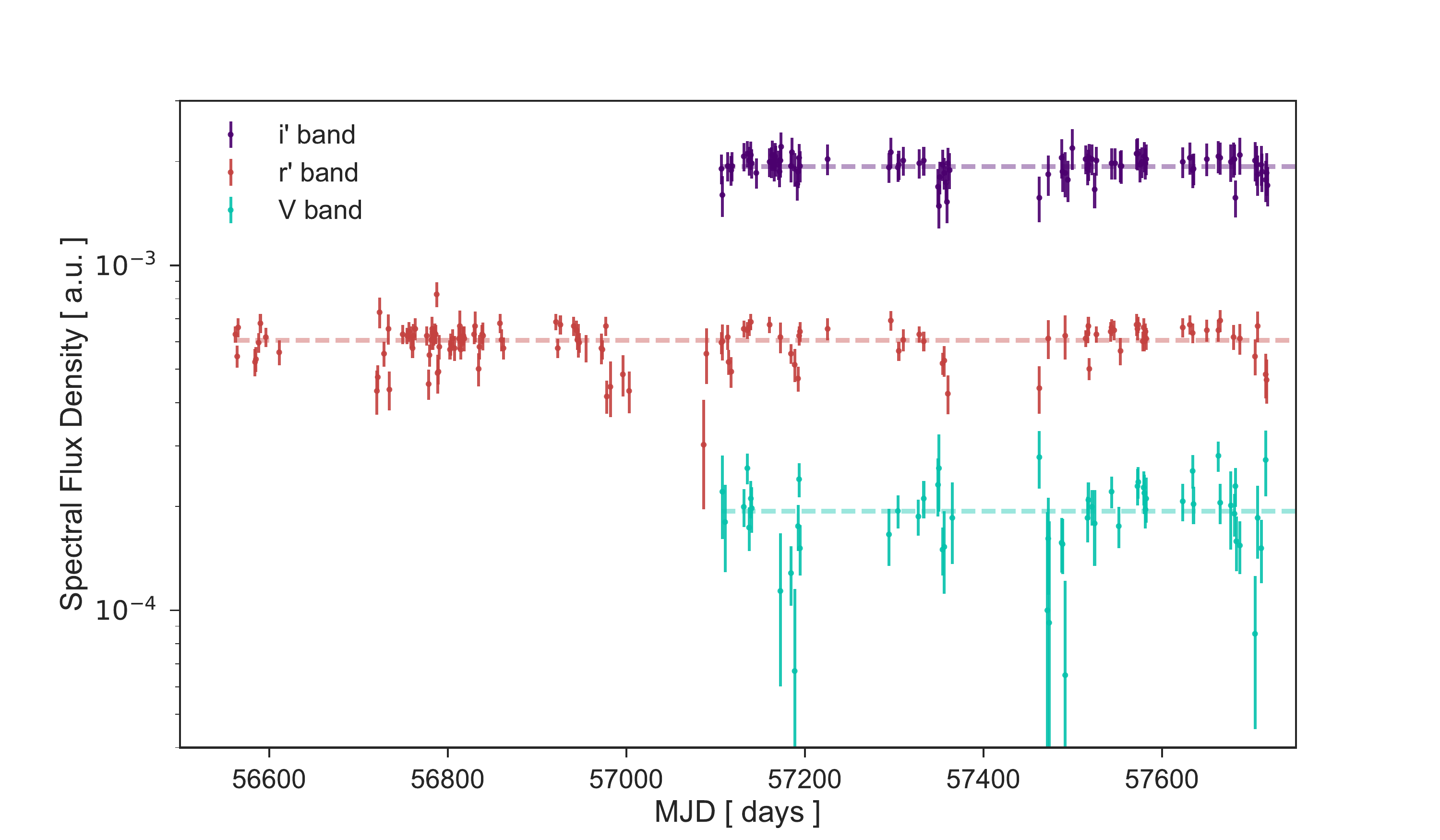}
\caption{Light curves from FLWO 48$''$ observations with Harris V (teal), SDSS r$'$ (red) and SDSS i$'$ (purple) filters. The spectral flux densities are given in arbitrary units (a.u.). The dashed lines show the average flux for each of the light curves.\label{fig:optical_lightcurves}}
\label{fig:48lightcurve}
\end{figure}

The FLWO 48$''$ (1.2~m) telescope is located on Mt. Hopkins in southern Arizona. As part of a long-term, multi-blazar observing program, optical photometry of HESS J1943+213 was obtained between 2013 September 27 and 2017 March 14 in SDSS r$'$ filter and between 2015 March 25 and 2017 March 14 in Harris V and SDSS i$'$ filters. The data reduction was performed using standard IDL tools. The magnitude zero-point was determined for each image by comparison to cataloged stars in order to derive the magnitude for HESS J1943+213 and a reference star located in the same field of view and only a few arcseconds from the source. The observations were not corrected for Galactic extinction, since extinction estimates at low Galactic latitudes are highly unreliable. Deviations of the reference star magnitude from the mean are used to reject observations from nights with poor weather. 
Magnitudes are converted to spectral flux densities assuming an AB magnitude system. The resulting light curves in the FLWO 48$''$ Harris V and SDSS r$'$ and i$'$ filters for HESS J1943+213 are presented in Figure~\ref{fig:48lightcurve}. For brighter, non-variable objects, such as the reference star used in this analysis, the distribution of measured magnitudes is consistent with the calculated statistical errors. We find, however, that for fainter objects with magnitude similar to HESS J1943+213, there is scatter in the measurements exceeding the calculated errors, indicating a dominant source of systematic error that remains to be identified. Hence, the FLWO 48$''$ data are not used for a variability search. Average fluxes are derived for each band and included in the source SED in Section~\ref{sec:sed}.

\subsection{Multi-frequency VLBA Observations in 2015 and 2016}
\label{sec:vlba}

The radio counterpart of HESS J1943+213 was observed by the authors using the VLBA. Observations took place over two epochs on 2015 August 11 (Project ID: BS246) and on 2016 August 08 (Project ID: BS253).

The 2015 observations (epoch I) were part of a request to follow up on the initial EVN detection and characterization of the source \citep{gabanyi13} by using four VLBA frequency bands (1.6 GHz, 4.3 GHz, 7.6 GHz, and 15 GHz). The epoch II observations taken in 2016 aimed to obtain deeper exposures of the source in C band (split into 4.3 GHz and 7.6 GHz bands) in order to characterize extended structures and measure polarization. The observations were targeted at the position reported from the EVN detection: $\alpha$ = 19$^{h}$43$^{m}$56$^{s}$2372, $\delta$ = 21$^{\circ}$18$'$23$''$402. All 10 VLBA antennae participated in both sets of observations. The total length of the 2015 observations was 4 hours, which included exposures on a phase calibrator source, J1946+2300, and a bandpass calibrator source, 3C 345. The 2016 observations totaled 8 hours and included exposures on the same phase calibrator and bandpass calibrator sources, as well as an astrometric check source, J1935+2031, and a polarized calibrator, 3C380.

\begin{figure}[ht]
\centering
\plotone{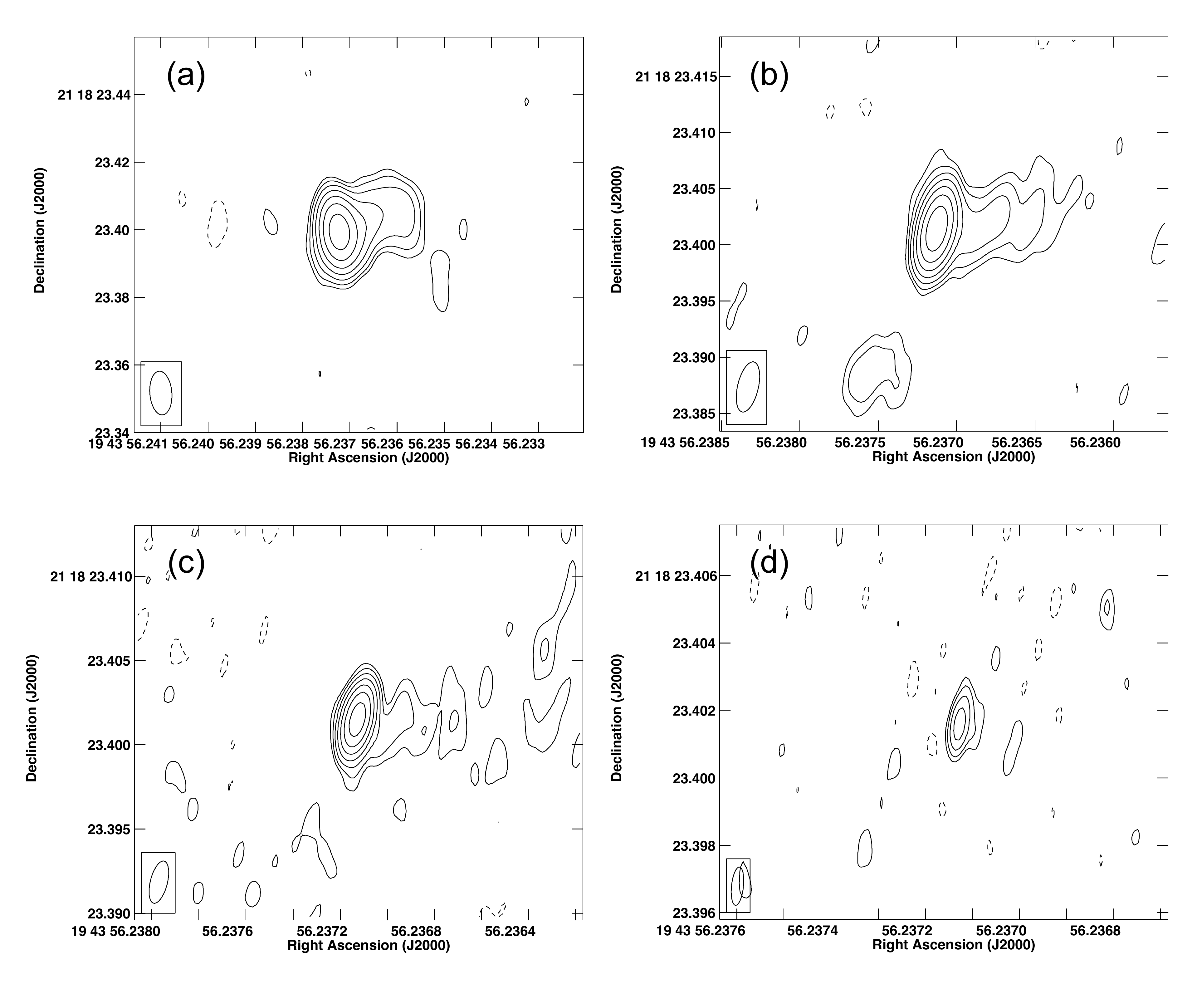}
\caption{Contour images of HESS J1943+213 with VLBA (a) 1.6~GHz, (b) 4.3~GHz, (c) 7.6~GHz, and (d) 15~GHz bands. Contour levels plotted above 1\% of the peak image intensity: $-$1, 1, 2, 4, 8, 16, 32, 64 (1.6~GHz); $-$2, $-$1, 1, 2, 4, 8, 16, 32, 64 (4.3~GHz and 7.6~GHz); $-$8, 8, 16, 32, 64 (15~GHz). Negative levels are shown with dashed lines.}
\label{VLBA_images}
\end{figure}

\begin{figure}[ht]
\epsscale{1.15}
\centering
\plottwo{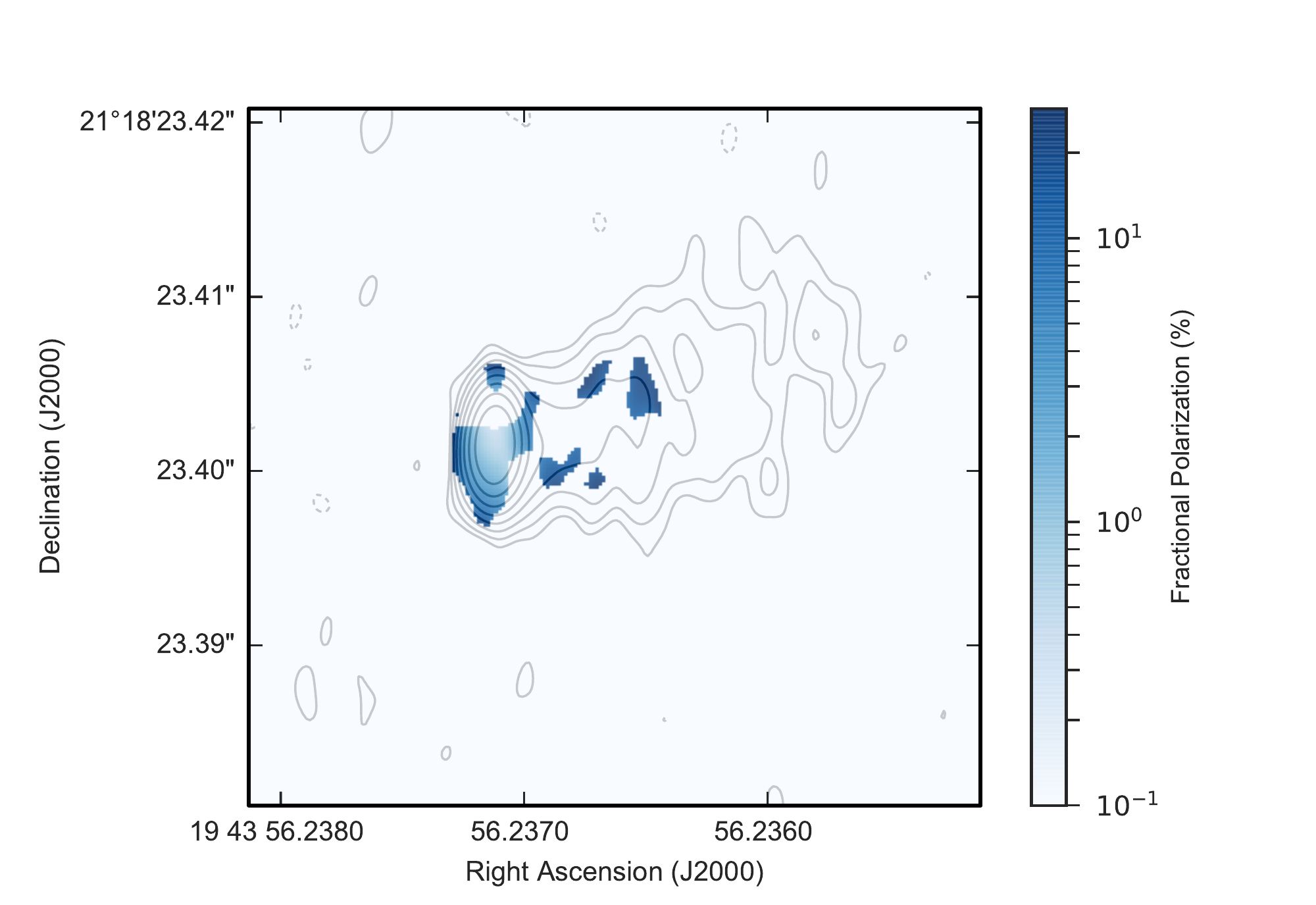}{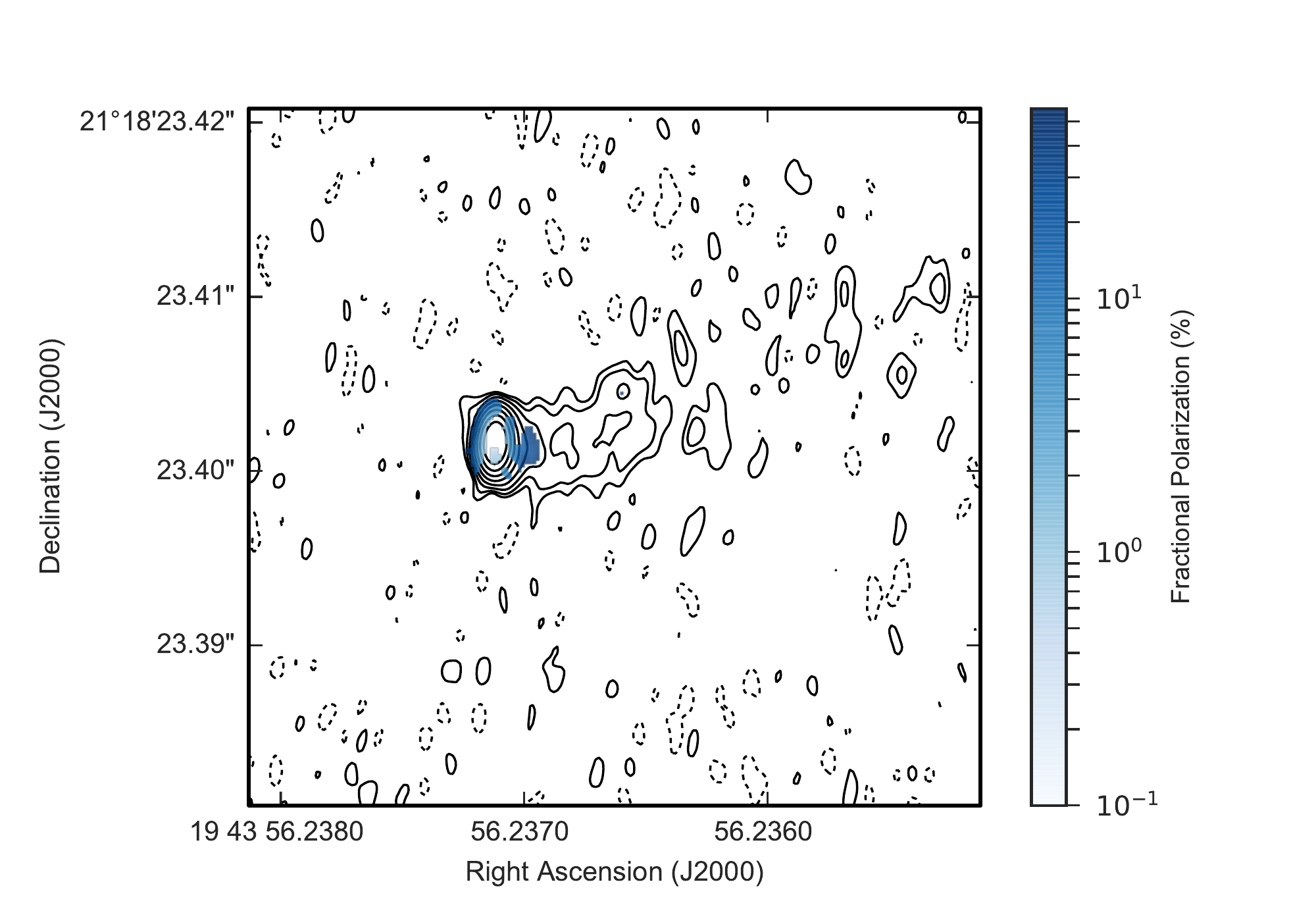}
\caption{Epoch II VLBA 4.3 GHz (left) and 7.6 GHz (right) images of HESS J1943+213 radio counterpart. The contours represent the $-1$ (dashed), 1, 2, 4, 8, 16, 32, 64, 128 levels above 0.3\% of the peak image intensity. The fractional polarization is illustrated with the color maps.}
\label{VLBA_pol}
\end{figure}

The NRAO Astronomical Image Processing System\footnote{http://www.aips.nrao.edu} (AIPS)~\citep{vanMoorsel96} is used to reduce and calibrate the VLBA data for HESS J1943+213. Images from the 2015 observations were produced for each band and are displayed in Figure~\ref{VLBA_images}. There is clear evidence for extended, jet-like emission in the 1.6 GHz, 4.3 GHz, and 7.6 GHz images. This is the first detection of the extended milliarcsecond-scale structure from the HESS J1943+213 counterpart in 4.3 GHz and 7.6 GHz bands, allowing multi-frequency exploration of its properties in VLBI. A similar core-jet structure has been previously detected with deep 1.6~GHz band observations with EVN \citep{akiyama16}. Images from the 2016 VLBA observations in 4.3 GHz and 7.6 GHz bands shown in Figure~\ref{VLBA_pol} reveal the source structure in more detail and provide fractional polarization measurements. The 1--3 \% polarization of the core in both bands detected for the source is consistent with polarization levels seen in other $\gamma$-ray blazars~\citep{linford12}. Quantitative results from the analysis of the two epochs of VLBA observations are provided in Table~\ref{VLBA}.

\floattable
\begin{deluxetable}{lccccl}
\tabletypesize{\scriptsize}
\tablecaption{ \small Measurements of HESS J1943+213 properties from phase and amplitude self-calibrated VLBA images.\label{VLBA}}
\tablewidth{0pt}
\tablehead{
\colhead{Band} & \colhead{Peak Intensity} & \colhead{Image Noise} & \colhead{Spectral Flux Density} & \colhead{Core Major / Minor Axis$^{\text{a}}$} & \colhead{T$_{B}$} \\
& \colhead{[mJy/beam]} & \colhead{[mJy/beam]} & \colhead{[mJy]} & \colhead{[mas]} & \colhead{[K]}
}
\startdata
\hline
\hline
\textbf{2015}&&&&\\
\hline
1.6 GHz & 18.564 & 0.073 & 23.31 $\pm$ 0.15 & 4.3 / 2.0 & $>$ 1.2 $\times$ 10$^{9}$  \\
4.3 GHz & 15.252 & 0.065 & 16.25 $\pm$ 0.12 & 0.68 / 0.24 & $>$ 2.2 $\times$ 10$^{9}$  \\
7.6 GHz & 15.032 & 0.071 & 16.23 $\pm$ 0.13 & 0.41 / 0.30 & $>$ 1.7 $\times$ 10$^{9}$  \\
$^{\text{b}}$15 GHz  & 8.1059 & 0.16 & 10.51 $\pm$ 0.32 & N/A / N/A & N/A  \\
\hline
\hline
\textbf{2016}&&&&\\
\hline
4.3 GHz & 18.388 & 0.020 & 17.12 $\pm$ 0.047 & 0.85 / 0.71 & $>$ 2.2 $\times$ 10$^{9}$  \\
7.6 GHz & 14.877 & 0.025 & 20.39 $\pm$ 0.041 & 0.57 / 0.47 & $>$ 1.5 $\times$ 10$^{9}$  \\
\hline
\enddata
\tablenotetext{\text{a}}{Sizes are reported after deconvolving by the beam size.}
\tablenotetext{\text{b}}{Values are from phase-only self-calibrated images.}
\end{deluxetable}

The core brightness temperature (T$_{B}$) of the HESS J1943+213 counterpart is estimated using images from all bands except for the 15 GHz band from 2015 observations, where the sensitivity was too low for phase and amplitude self-calibration. Lower limits to T$_{B}$ are derived due to a partially resolved core and the possibility of interstellar scattering, resulting in T$_{B}$ $>$~1.2$\times$10$^{9}$~K in the most conservative case, measured with the 1.6 GHz image from the 2015 dataset. The brightness temperature values are well within the range for blazars. 
We do not confirm the significantly lower brightness temperature measurement of T$_{B}$ = 7.7$\times$10$^{7}$~K, which was based on the EVN 1.6 GHz image \citep{gabanyi13} and which has subsequently been reanalyzed showing higher brightness temperature, T$_{B}$ $>$ 1.8$\times$10$^{9}$ K \citep{akiyama16}. 

\begin{figure}[ht]
\plotone{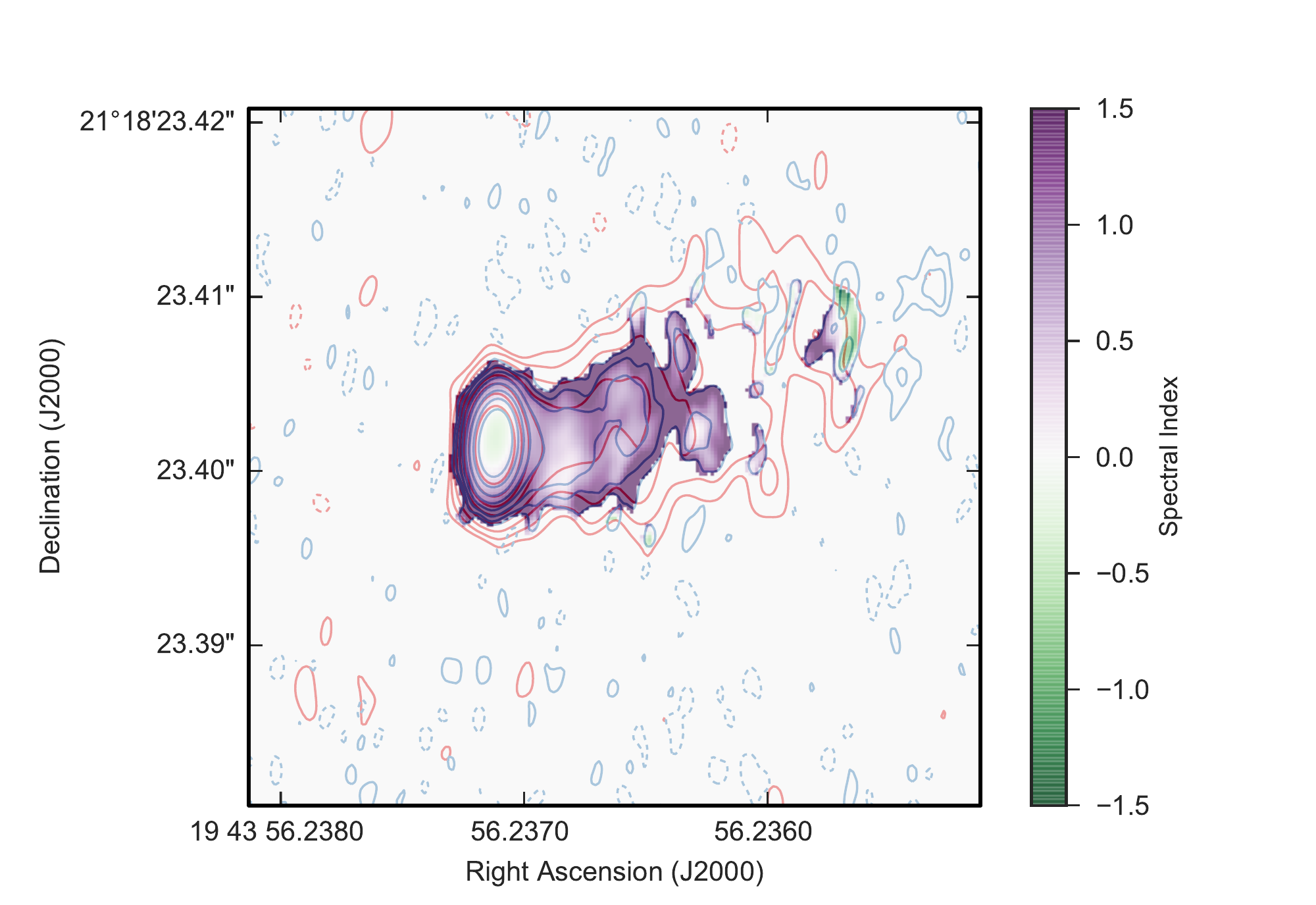}
\caption{Spectral map of the core-jet structure of HESS J1943+213 radio counterpart made from epoch II VLBA 4.3 GHz and 7.6 GHz (degraded to 4.3 GHz resolution) band images, with contours for 4.3 GHz (red) and 7.6 GHz (blue) representing $-1$ (dotted), 1, 2, 4, 8, 16, 32, 64, 128 levels above 0.3\% of the peak image intensity.}
\label{VLBA_index}
\end{figure}

Using the 2015 observations and assuming a power law function of the form S $\propto$ $\nu^{-\alpha}$, where S is the spectral flux density, a spectral index $\alpha_{\text{core}}$ = 0.3 $\pm$ 0.06 is measured for the HESS J1943+213 core, determined from all 4 bands, and an index 
$\alpha_{\text{jet}}$ = 1.1 $\pm$ 0.4 for the extended emission is calculated over 9.3 square milliarcseconds based on the 4.3 GHz and 7.6 GHz images. With the deeper 2016 dataset, we construct a spectral index map and present it in Figure~\ref{VLBA_index}. In order to construct the map, the 7.6 GHz image is convolved with a larger beam size to match the resolution of the 4.3 GHz image. There is a visible discrepancy between $\alpha_{\text{core}}$ = 0.3 $\pm$ 0.06 determined from epoch I observations using the 1.6, 4.3, 7.6, and 15 GHz images and the values shown in the spectral index map that are near $\alpha = -0.3$ at the center of the core. This is largely a result of differing resolutions for the four bands involved in the spectral index calculation for epoch I data. The epoch II spectral index map, which uses images with matched resolutions is more robust and does not suffer from this issue.

Measurements of the spectral indices of the core and the extended structures from both epochs are consistent with reported values for blazar cores and jets from the MOJAVE sample \citep{hovatta14}. The only available radio spectral index measurements of the core of this source come from e-MERLIN observations, which find an index of 0.03 $\pm$ 0.03~\citep{straal16}. The e-MERLIN observations do not resolve the source, however, and the core spectral index calculation using these observations is affected by emission from extended structures.

Comparisons between EVN and VLBA 1.6 GHz results show apparent changes in the core flux density. \citet{gabanyi13} measured 31 $\pm$ 3 mJy for the source flux density with EVN, while the flux density measurement from our VLBA image is 23.6 $\pm$ 0.2 mJy. A similar change in the source flux density was reported by \citet{straal16} between the 2011 EVN 1.6 GHz result and a lower resolution e-MERLIN 1.5 GHz detection in 2013 of the source with 22.2 $\pm$ 0.7 mJy, leading to the first variability claim for HESS J1943+213 or its counterparts in any band. As the core is not fully resolved in 1.6 GHz, however, and different configurations were used for the VLBA, EVN, and e-MERLIN observations, the claims for variability are not definitive.

e-MERLIN observations in the 5 GHz band were also obtained by \citet{straal16}, resulting in a 22.4 $\pm$ 0.3 mJy flux density, which is significantly higher than the VLBA 4.3 GHz measurements of $\sim$16.2 mJy in 2015 and 17.1 mJy in 2016. The latter discrepancy could be explained by a change in the source flux density, but more likely by differences in \textit{uv} coverage between VLBA and e-MERLIN observations and the inclusion of the jet feature in the core flux density measurement with e-MERLIN.

Despite the strong arguments for classifying HESS J1943+213 as an EHBL, a measurement of the proper motion can be a definitive discriminator between Galactic and extra-Galactic origin for the source.
We attempt two sets of proper motion measurements of the HESS J1943+213 VLBI counterpart. In the first case, we compare the position measurement from our pure phase-referenced (no self-calibration) VLBA 15 GHz image from 2015 to the position reported from the \citet{gabanyi13} EVN detection, and find a change in position of 1.1 mas. This is consistent with zero, given the $\sim$2.5 mas uncertainty in the position measurements, which is largely dominated by the uncertainty in the EVN position. Using this uncertainty and the $\sim$4.3-year time difference between the two observations, an upper limit of 47 km/s is calculated for the transverse velocity of the source at 17~kpc -- the assumed distance if the source is a Galactic PWN~\citep{hess11,gabanyi13}. 

In the second case, we compare the positions from our 2015 and 2016 VLBA observations using the phase-referenced 4.3 GHz images almost exactly a year apart. The 2015 VLBA observations lacked a reference source for determining an absolute source position. Hence, we make a conservative estimate of the uncertainty in the position measurements. The uncorrected atmosphere and ionosphere will contribute $<$~0.1 mas to the positional uncertainty~\citep[e.g.,][]{hachisuka15}. The high image signal-to-noise ratio ($>$ 1000:1) ensures that the statistical
contribution to the uncertainty is well below 0.01 mas and is effectively negligible. The biggest source of uncertainty is the effect of source structure on the measurements.  
Source structure can complicate position measurement (1) through unmatched \textit{uv} coverage of the two epochs leading to slight differences in reconstruction of the source and (2) through evolution of the source itself. The maximum contribution from source structure to the uncertainty in the position measurement can be estimated from the ratio of the core brightness to the
local structure (in Jy/Beam). At worst, this ratio is 30:1 and the uncertainty from source structure is 1/30 of a beam width (3 $\times$ 1.5 mas), i.e. $\sim$0.1 mas. Thus, a conservative estimate of 0.1 mas is adopted as the uncertainty in the position measurements between the two VLBA observations.

The source position in the 4.3 GHz images is measured by fitting a Gaussian function to the core. The differences in the position measurements are 0.08 $\pm$ 0.04 mas, where the uncertainty is from the fit statistics alone. The measured difference in the position between the two epochs is smaller than the uncertainty in the position determination. We again set an upper limit on the transverse velocity. In this case, with the same source distance assumption of 17~kpc, the upper limit on the transverse velocity is 8 km/s.

The upper limits on the transverse velocity, especially the 8 km/s limit from the two epochs of VLBA observations, are much lower than typical Galactic pulsar velocities -- between 100 and 300 km/s -- obtained from VLBI proper motion measurements~\citep{brisken03}. These velocity limits break the assumption that the source is located within the Galaxy and effectively rule out a Galactic origin for the source.

\section{Results and Discussion}
\label{sec:interp}

Recent publications have strongly argued that HESS J1943+213 is a blazar and results presented above, especially the VLBA measurements, firmly support and solidify this scenario. Based on the location of the synchrotron peak, HESS J1943+213 is characterized as an EHBL, a blazar subclass with very few detected members. HE $\gamma$-ray blazars behind the Galactic Plane have been previously identified with \textit{Fermi}-LAT \citep{kara12}, but HESS J1943+213 is the first such blazar also seen in VHE $\gamma$-rays.

\subsection{Redshift Constraints from Gamma-ray Spectra}
\label{sec:redshift}

The improved detection of the inverse-Compton peak with \textit{Fermi}-LAT and VERITAS and the resulting higher-statistics spectra are used to set more robust upper limits on the redshift of the source. We use the same procedure for redshift estimation of HESS J1943+213 as \citet{peter14}. In this method, the \textit{Fermi}-LAT power-law spectrum is assumed to be the proxy for the intrinsic $\gamma$-ray spectrum and downwards deviations from a power-law shape are attributed to absorption effects from pair-production interactions between $\gamma$-rays and EBL photons. 

\begin{figure*}[ht]
\plotone{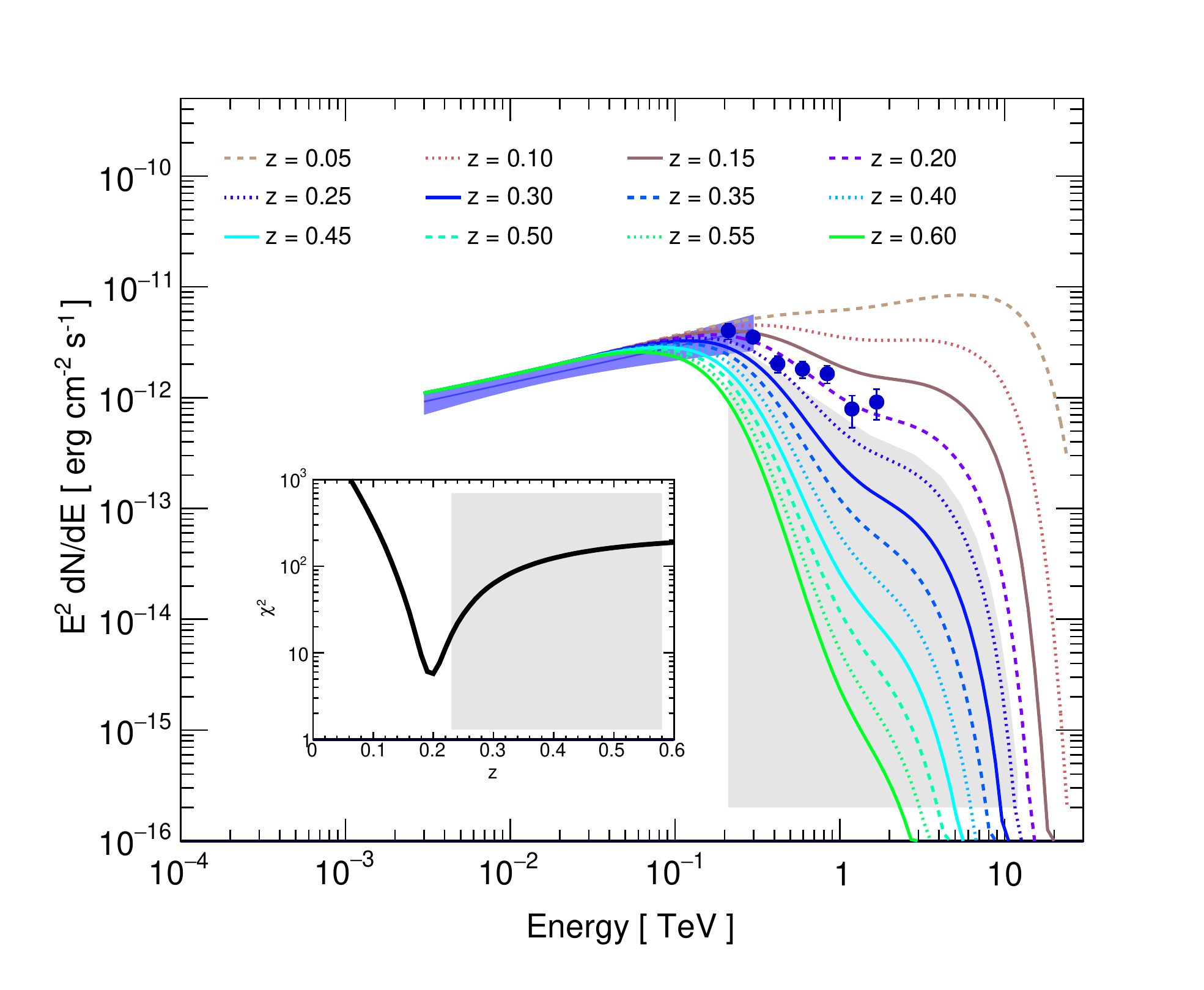}
\caption{VERITAS observed spectrum (blue points) fit to upper bound of the \textit{Fermi}-LAT spectrum absorbed by EBL for redshift values ranging from 0 to 0.6. The inset figure shows the $\chi^2$ distribution with redshift of the VERITAS spectrum fit to the EBL-absorbed extrapolations of the \textit{Fermi}-LAT upper bound. The gray-shaded areas show the 95\% rejection regions.}
\label{fig:redshift}
\end{figure*}

Assuming a model of the EBL by \citet{franceschini08}, the 68\% upper bound of the \textit{Fermi}-LAT spectrum is extrapolated into the VHE regime and absorbed for a range of redshift values. The upper bound of the \textit{Fermi}-LAT spectrum is used in order for the upper limit on the redshift to be conservative. The $\chi$$^{2}$ value is calculated from the extrapolated \textit{Fermi}-LAT upper bound and the VERITAS spectral points for each redshift value. The observed VERITAS spectrum along with the extrapolated \textit{Fermi}-LAT upper bounds for a range of redshifts are shown in Figure~\ref{fig:redshift}. The figure also includes the resulting $\chi$$^{2}$ distribution, which shows a minimum $\chi$$^{2}$ value at z $\sim$ 0.20. The 95$\%$ upper limit on the redshift derived from the $\chi$$^{2}$ distribution is z $<$ 0.23, which is significantly more constraining than the existing z $<$ 0.45 95$\%$ upper limit from~\citet{peter14}.

\subsection{Search for Flux Variability in X-rays and $\gamma$-rays}
\label{sec:var}

So far, the only claim of variability from HESS J1943+213 and its identified multi-wavelength counterparts comes from measurements of different flux densities of the radio core on milliarcsecond scales using VLBI observations \citep{akiyama16}.

Light curves from VERITAS and \textit{Swift}-XRT are presented in Figure~\ref{fig:veritas_lightcurve} and inset of Figure~\ref{fig:xrt_spectra}. A simple $\chi^{2}$ fit of a constant line to each light curve is used to test for flux variability. In addition, the fractional root mean square variability amplitude \citep{edelson90,rodriguez97} is calculated for each light curve, defined as F$_{\text{var}}$ = $\sqrt{\frac{\sigma^{2} - \delta^{2}}{\langle f \rangle^{2}}}$, where $\sigma^2$ is the variance of the fluxes, $\delta^2$ is the mean square uncertainty of the fluxes, and $\langle$f$\rangle$ is the mean flux. The uncertainty in F$_{\text{var}}$ is given by Equation B2 in~\citet{vaughan03}.

The long-term VERITAS light curve is stable, with F$_{\text{var}}$ = 0.23 $\pm$ 0.37 and $\chi^2$ / NDF = 15.6 / 12 corresponding to a p-value of 0.21 for a constant flux. There is no statistically significant evidence for variability in the \textit{Swift}-XRT light curve composed of three observations with F$_{\text{var}}$ = 0.007 $\pm$ 0.003 and $\chi^2$ / NDF = 12.0 / 2 corresponding to a p-value of 0.003. In addition, there is no evidence of variability within individual XRT observations. HESS J1943+213 remains one of the most stable VHE-detected blazars.

\subsection{Modeling the HESS J1943+213 Spectral Energy Distribution}
\label{sec:sed}

\subsubsection{SED Construction and Assumptions}

\floattable
\begin{deluxetable}{ccc}
\tabletypesize{\scriptsize}
\tablecaption{ \small Parameters of the SSC models for the HESS J1943+213 broadband SED.\label{tab:sedparams}}
\tablewidth{0pt}
\tablehead{
& \colhead{GENERAL TRANSFORMATION PARAMETERS} &
}
\startdata
0.16 & redshift &\\
71 & Hubble Constant &  km s$^{-1}$ Mpc$^{-1}$ \\
2.0 & Angle to the Line of Sight & degrees \\
\hline
\hline
& BLOB PARAMETERS &\\
\hline
26 & Doppler factor, $\delta$ & \\
3.8$\times$10$^{3}$ & Particle density, K & cm$^{-3}$ \\
1.9 & First slope of particle energy spectrum, $\alpha_{1}$ & \\ 
3.0 & Second slope of particle energy spectrum, $\alpha_{2}$ &\\
8.0$\times$10$^{3}$ & Minimum electron energy Lorentz factor, $\gamma_{\text{min}}$ & \\
5.0$\times$10$^{6}$ & Maximum electron energy Lorentz factor, $\gamma_{\text{max}}$ & \\
2.0$\times$10$^{5}$ & Break in electron energy spectrum, $\gamma_{\text{b}}$ & \\
0.1 & Magnetic field, B & G \\
3.0$\times$10$^{15}$ & Radius of emitting region, R& cm \\
\hline
\hline
& JET PARAMETERS (1ST SLICE) &\\
\hline
9 & Doppler factor & \\
6.0$\times$10$^{2}$ & Particle density & cm$^{-3}$ \\
2.0 & Slope of particle energy spectrum & \\ 
2.0$\times$10$^{2}$ & Minimum electron energy Lorentz factor & \\
1.1$\times$10$^{4}$ & Maximum electron energy Lorentz factor & \\
0.2 & Initial magnetic field & G \\
1.2$\times$10$^{16}$ & Inner radius (host galaxy frame) & cm \\
\hline
& JET PARAMETERS (GLOBAL) &\\
\hline
8.6 & Jet length (host galaxy frame) & pc\\
2.3 & Half opening angle of jet (host galaxy frame) & deg\\
50 & Number of slices & \\
0.3 & Minimum \textit{blob} -- black hole distance (host galaxy frame) & pc \\
\enddata
\end{deluxetable}

\begin{figure}[ht]
\plotone{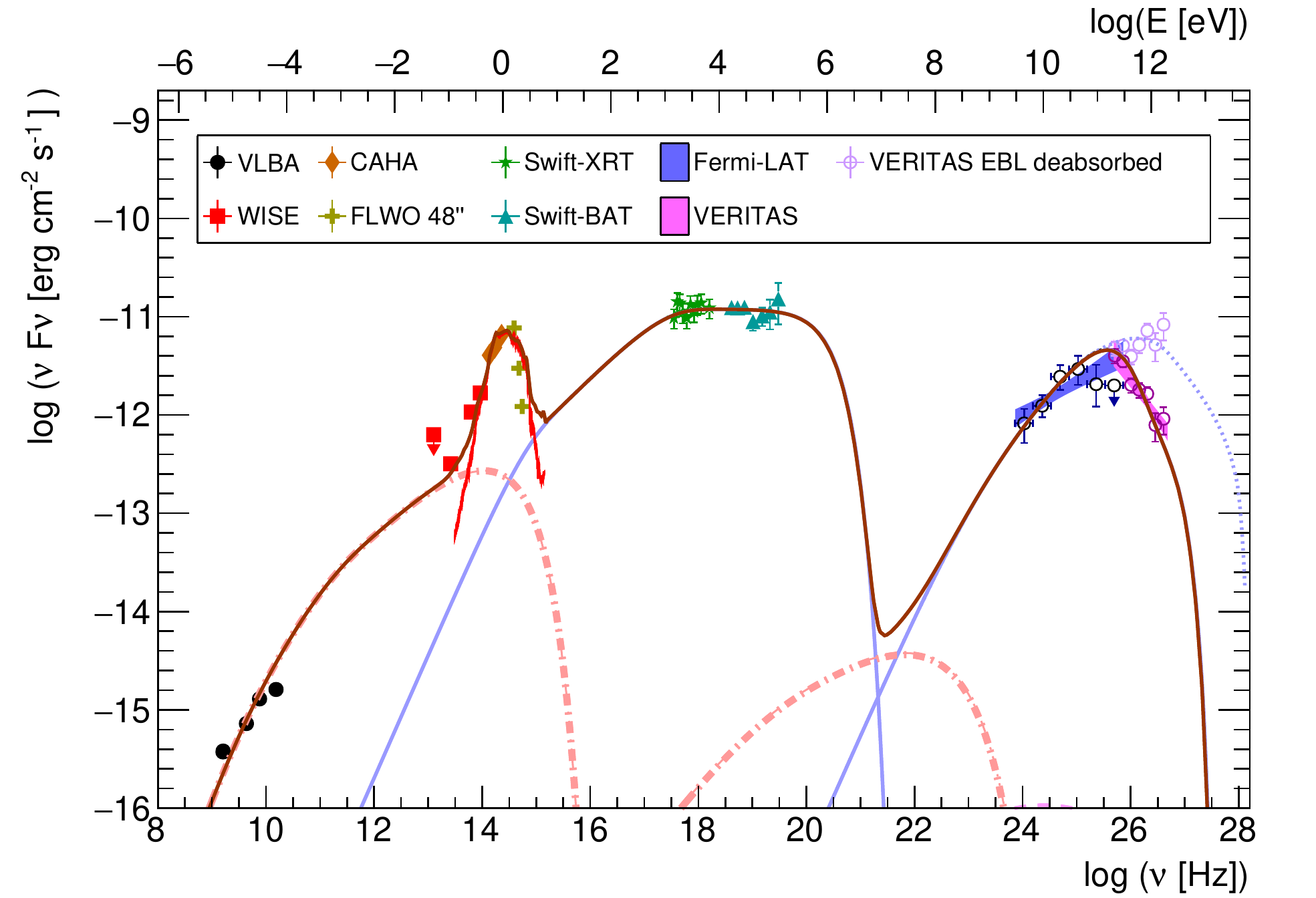}
\caption{SED of HESS J1943+213, including the SSC model with components for a \textit{blob} of relativistic particles (solid light blue curves) and a larger \textit{jet} (dash-dotted red curves). The model for the host galaxy emission is shown in solid red, while the brown curve gives the summed emission from all model components. The flux points include data from VLA 1.4 GHz, VLBA 1.6, 4.3, 7.6, 15~GHz, WISE, CAHA 3.5~m~\citep{peter14}, FLWO 48$''$, \textit{Swift}-XRT and \textit{Swift}-BAT, \textit{Fermi}-LAT and VERITAS. The assumed source redshift is z = 0.16. The EBL model from~\citet{franceschini08} is used to deabsorb the VERITAS points.}
\label{fig:sed}
\end{figure}

The time-averaged broadband SED of HESS J1943+213 is displayed in Figure~\ref{fig:sed}. The SED is assembled using data analyzed in this work from VERITAS, \textit{Fermi}-LAT, \textit{Swift}-XRT, FLWO 48$''$, and VLBA and archival SED points in the hard X-rays from the \textit{Swift} Burst Alert Telescope (BAT) 70-month survey~\citep{70bat} and in the infrared from the Wide-field Infrared Survey Explorer (WISE) \citep{wise} all-sky survey and Calar Alto Astronomical Observatory (CAHA) 3.5~m Telescope observations~\citep{peter14}. As noted in previous publications~\citep{hess11,tanaka14,peter14}, there is no detected cutoff in hard X-rays up to an energy of 195 keV, supporting the classification of the source as an EHBL. With the improved HE and VHE spectra of the source from VERITAS and 8-year \textit{Fermi}-LAT observations respectively, the inverse-Compton peak of the SED is well constrained. 

For the purposes of modeling the HESS J1943+213 SED, we assume a source redshift of z = 0.16. This redshift value is derived by repeating the redshift estimation procedure in Section~\ref{sec:redshift} with the nominal Fermi-LAT spectrum instead of the Fermi-LAT upper bound and selecting the redshift where the $\chi^{2}$ distribution reachers a minimum. The estimate assumes the Fermi-LAT spectrum does not deviate from a power-law function; hence, it is likely to be an overestimate if any downward curvature is present in the intrinsic source spectrum.

\clearpage
\subsubsection{Model Description and Constraints}

We model the HESS J1943+213 SED with a two-zone SSC model described by a homogenous, compact \textit{blob} within a conical wider \textit{jet}. The model is a two-flow representation in which there is a highly Doppler boosted inner jet region (\textit{blob}) embedded in a wider, conical structure with a lower Lorentz factor (\textit{jet})~\citep[][based on~\citet{katarzynski01}]{hervet15}. This type of model is supported by theoretical jet approaches \citep[e.g.,][]{sol89}, jet production mechanisms~\citep{blandford77,blandford82}, jet launching simulations~\citep{ferreira06}, and radio VLBI imaging~\citep{mertens16}.
Our model is similar to the two-zone models used by~\citet{ghisellini05} and~\citet{shukla16}, but with notable distinctions. The~\citet{ghisellini05} model used a spine-layer structure, with two cylindrical components consisting of a faster moving, higher Lorentz factor inner spine, embedded in a slower, lower Lorentz factor outer layer. The~\citet{shukla16} model built on the spine-layer model with the addition of another SSC component near the base of the jet to represent a harder, variable emission. Unlike these models, the~\citet{hervet15} model used here employs a conical geometry for the wider, lower Lorentz factor emission region (\textit{jet}), with a spherical, higher Lorentz factor emission region (\textit{blob}) imbedded within.

The \textit{blob} SSC component assumes an emission region composed of relativistic particles, containing tangled magnetic fields, and moving towards Earth with a Doppler factor $\delta$. The particle population is described by a broken power-law function with indices $\alpha_{1}$ and $\alpha_{2}$, and minimum, maximum, and break energies ($\gamma_{\text{min}}$, $\gamma_{\text{max}}$, and $\gamma_{\text{b}}$). The size of the \textit{blob} is chosen to best represent the multi-wavelength SED while staying close to equipartition and within the standard range of sizes for blazar models. The EBL model from~\citet{franceschini08} is used to calculate the attenuation of $\gamma$-rays due to pair-production interactions for the SED model.

The SED model includes a near-IR and optical emission component from the host galaxy~\citep[using a PEGASE 2 template from][]{fioc99}, which is characterized as a giant elliptical galaxy with a lower limit on the mass of 2.0 $\times$ 10$^{11}$ M$_\odot$.
The host galaxy model fits the SED data well and is the preferred description for the near-IR SED points. A non-thermal origin for this emission is unlikely and would be impossible to accommodate within the broadband SSC model. An alternative explanation for the near-IR excess is that the emission comes from a dust torus around the central black hole. This is also unlikely, however, since strong dust torus emission is not expected for HBLs and EHBLs \citep[e.g.,][]{meyer11}. In addition, a bright torus would induce an external-Compton signature in $\gamma$-ray energies, which we do not observe.

The VLBA and the mid-IR points are modeled with SSC emission from a stratified, conical \textit{jet}. The \textit{jet} model is represented as a cone discretized into 50 cylindrical slices, with increasing slice volumes at larger distances from the jet basis. Only the radius of the first slice, the jet length, and the jet opening angle are used as priors to define the jet structure. The thickness of each slice is increased logarithmically with the distance to the jet base to attain roughly the same number of particles in each slice. The radius of each slice is determined by the distance of the slice to the base and the jet opening angle. The \textit{jet} speed is assumed to be constant along its propagation. The magnetic field decreases inversely with the distance to the first slice following radio measurements by~\citet{sullivan09}. \textit{Jet} parameters such as the magnetic field strength and particle density are provided for the first cylindrical slice in Table~\ref{tab:sedparams}. These parameters are calculated for all other slices following an adiabatic expansion evolution. Absorption and emission coefficients are calculated for each slice. 

The synchrotron self-Compton radiation from each \textit{jet} slice is used to calculated the total emission from the \textit{jet} after taking into account radiation transfer through all slices in the direction of the \textit{jet} propagation. The radiation transfer of the \textit{blob} emission through the slices between the \textit{blob} and the observer are also calculated. In addition, the external inverse-Compton (EIC) interaction between the \textit{blob} and the \textit{jet} components are computed. The \textit{jet} radiation transfer at the \textit{blob} position is calculated in the jet frame. This radiation field is converted into the blob frame to calculate the EIC radiation and the resulting EIC radiation is transferred through the jet slices in the direction of the observer. Inverse-Compton radiation between the \textit{jet} particles and the radiation field from the \textit{blob} are assumed to be negligible. The full details of the jet model can be found in~\citet{hervet15}.

The following constraints are applied to simulate the \textit{jet} emission: (1) \textit{blob} velocity is greater than \textit{jet} velocity, (2) radius of the \textit{jet} is larger than the radius of the \textit{blob}, (3) \textit{blob} $\gamma_{max}$ is greater than \textit{jet} $\gamma_{max}$, and (4) the \textit{jet} is very close to equipartition.

In addition, information from VLBA imaging of the radio counterpart presented in Section~\ref{sec:vlba} can place additional constraints on the \textit{jet} parameters. We use the size of the radio core measured from the VLBA 7.6 GHz image during Epoch 1, as it provides the most stringent constraints on \textit{jet} model parameters. The jet direction appears perpendicular to the core major axis. Assuming a conical jet basis, the core major axis gives the maximum diameter of the jet basis. The projected length of the jet basis (on the sky plane) is estimated with $\texttt{minor axis} - \frac{\texttt{major axis}}{2}$, which in turn gives an apparent jet half-opening angle of 65.2$^{\circ}$. With the usual value of the angle between the jet axis and the line-of-sight of 2$^{\circ}$, the intrinsic jet half-opening angle and the intrinsic radio core or jet basis length are estimated as 2.3$^{\circ}$ and 8.59~pc respectively.

\subsubsection{SED Modeling Results and Energetics Discussion}

This two-flow model is able to represent the SED very well, only slightly underestimating the lower energy VLBA point. The $\chi^{2}$/NDF goodness of fit values for the \textit{blob} model are $\chi^{2}_{\text{XRT}}$/NDF = 4.16/10, $\chi^{2}_{\text{BAT}}$/NDF = 3.38/7, $\chi^{2}_{\text{LAT}}$/NDF = 3.09/5, and $\chi^{2}$$_{\text{VERITAS}}$/NDF = 11.33/7 (note that NDF does not include the number of free parameters of the SED model). The parameters governing the wider \textit{jet} model are poorly constrained, given there is only one synchrotron slope. As such, the physical values and the energetics of this extended \textit{jet} emitting zone are highly model-dependent.

The full list of values of the SED model parameters can be found in Table~\ref{tab:sedparams}. The synchrotron peak is very broad and is located between 10$^{18}$~Hz and 10$^{20}$~Hz according to the model. The minimum variability timescale predicted by the SED model is $\sim$1.24 hours for the \textit{blob} and $\sim14.3$ hours for the \textit{jet}. These timescales are not contradicted by the lack of variability detection in X-rays and $\gamma$-rays, especially if the system is in a thermal equilibrium with constant particle injection.
The large \textit{blob} Doppler factor value of 26 used in the SED model leads to a low internal pair-production opacity up to the highest observed gamma-ray energies, with an optical depth of $\sim0.03$ at 1~TeV.

The \textit{blob} emission region is out of equipartition, with the energetics dominated by the kinetic energy of the particles and u$_{B}$/u$_{e}$ = 0.01, where u$_{e}$ is the energy density in the particles and u$_{B}$ is the energy density in the magnetic fields. The \textit{jet} emission region, on the other hand, is at equipartition. The gamma-ray peak of the observed SED does not show signs of extra broadening, implying that the EIC emission is effectively negligible for the source. If the \textit{blob} is too close to the central black hole, \textit{blob} particles will strongly interact with synchrotron radiation from the \textit{jet} basis and produce a strong EIC component. Thus, the lack of observed EIC emission can be used to place a lower limit on the distance between the \textit{blob} and the central supermassive black hole (SMBH). Using the jet parameters of the SED model, the first jet slice is located at a distance of 0.1~pc from the SMBH and the constraint on the \textit{blob}-SMBH distance is $\geq0.3$~pc, with the gamma-ray emission peak containing $\sim$5\% EIC and 95\% SSC radiation.

The choice of parameters for the presented model is based on a good representation of the multi-wavelength SED. Once this is achieved, an effort is made to stay close to equipartition and to reduce the total \textit{jet} energetics. Parameter degeneracies, intrinsic to SSC models, cannot be fully broken by this approach. With the strong observational constraints on the synchrotron and $\gamma$-ray peaks, however, significant changes to the presented parameters will require moving away from these modeling criteria.

Given the exact distance of the source is still uncertain, changing the assumed distance value will change the intrinsic power of the source; however, the effect on the energetics equilibria including equipartition will not be significant. A higher (lower) redshift value will yield a higher (lower) energy of the emitting particles, and imply a stronger (weaker) particle acceleration mechanism. We tested models with different source redshift assumptions (z = 0.1 -- 0.2) and obtained a range of likely parameters. The \textit{blob} parameters with most significant changes are the Doppler factor (18 -- 30), particle density (8$\times$10$^{2}$~cm$^{-3}$ -- 4$\times$10$^{3}$~cm$^{-3}$), and radius (3$\times$10$^{15}$~cm -- 7$\times$10$^{15}$~cm).

There have been two previous efforts to model the SED of this source. \citet{tanaka14} modeled the IR-to-$\gamma$-ray SED using a blackbody component for the host galaxy and an SSC component. Their SSC model includes magnetic fields with a strength of 0.78 mG, a Doppler factor of 70, and a single power-law electron distribution characterized by a spectral index of 3, $\gamma_{min}$ = 10$^{5}$, $\gamma_{max}$ = 3 $\times$ 10$^{7}$. The derived variability timescale in this case is $\sim$28 hours and the model is far from equipartition with u$_{B}$/u$_{e}$ = 0.001. The other SED model for the source comes from~\citet{peter14}, which represents the entire SED including the radio regime with emission from a single population of electrons and a blackbody component for the host galaxy. In this case, the magnetic field strength is 0.05 G, the electron population is described as an exponential cut-off power-law function with $\gamma_{min}$ = 1, $\gamma_{max}$ = 10$^{10}$ and the energetics are out of equipartition and dominated by kinetic energy of particles with u$_{B}$/u$_{e}$ = 0.08. 

In terms of energy requirements, our model is able to reproduce the SED with a significantly lower value for the $\gamma_{max}$ than~\citet{tanaka14} and~\citet{peter14} and a much lower value of the Doppler factor than~\citet{tanaka14}. The magnetic field strength in the~\citet{tanaka14} model is lower than ours; however, this results in an emission zone that is very far from equipartition. Overall, the SED model presented in this work is able to fit the data well using more standard parameters for HBLs. In addition, the more constraining $\gamma$-ray data makes our model more robust than previous attempts.

\subsection{The Role of UHECR Cascade Emission}
\label{sec:crpropa}

Despite the SSC scenario providing a good description for the HESS J1943+213 $\gamma$-ray emission, and in light of HESS J1943+213 being identified as an EHBL -- a promising class of objects for hadronic emission -- we investigate the possibility that the VHE $\gamma$-rays originate instead from electromagnetic cascades produced by interactions of UHECRs with background photon fields. To estimate the secondary $\gamma$-ray emission from such a scenario, we simulate the propagation of UHECRs and calculate all relevant interactions using publicly-available software, CRPropa3 ~\cite[for details of the software package see][]{crpropa3}. 

\floattable
\begin{deluxetable}{ccc}
\tabletypesize{\scriptsize}
\tablecaption{ \small Parameters used in modeling the $\gamma$-ray data with UHECR-induced cascade emission.\label{tab:crpropa}}
\tablewidth{0pt}
\tablehead{
& \colhead{GENERAL PARAMETERS} &
}
\startdata
124 -- 1387 & Source Distance & Mpc\\
71 & Hubble Constant &  km s$^{-1}$ Mpc$^{-1}$ \\
10 & Intergalactic magnetic field strength & fG\\
\hline
\hline
& COSMIC RAY PARAMETERS &\\
\hline
Protons only & Composition & \\
2.0 & Index of cosmic ray spectrum & \\
0.7 & Minimum cosmic ray energy & EeV \\ 
300 & Maximum cosmic ray energy & EeV\\
6.0 & Jet opening angle for cosmic rays & degrees \\
\enddata
\end{deluxetable}

\begin{figure}[]
\plotone{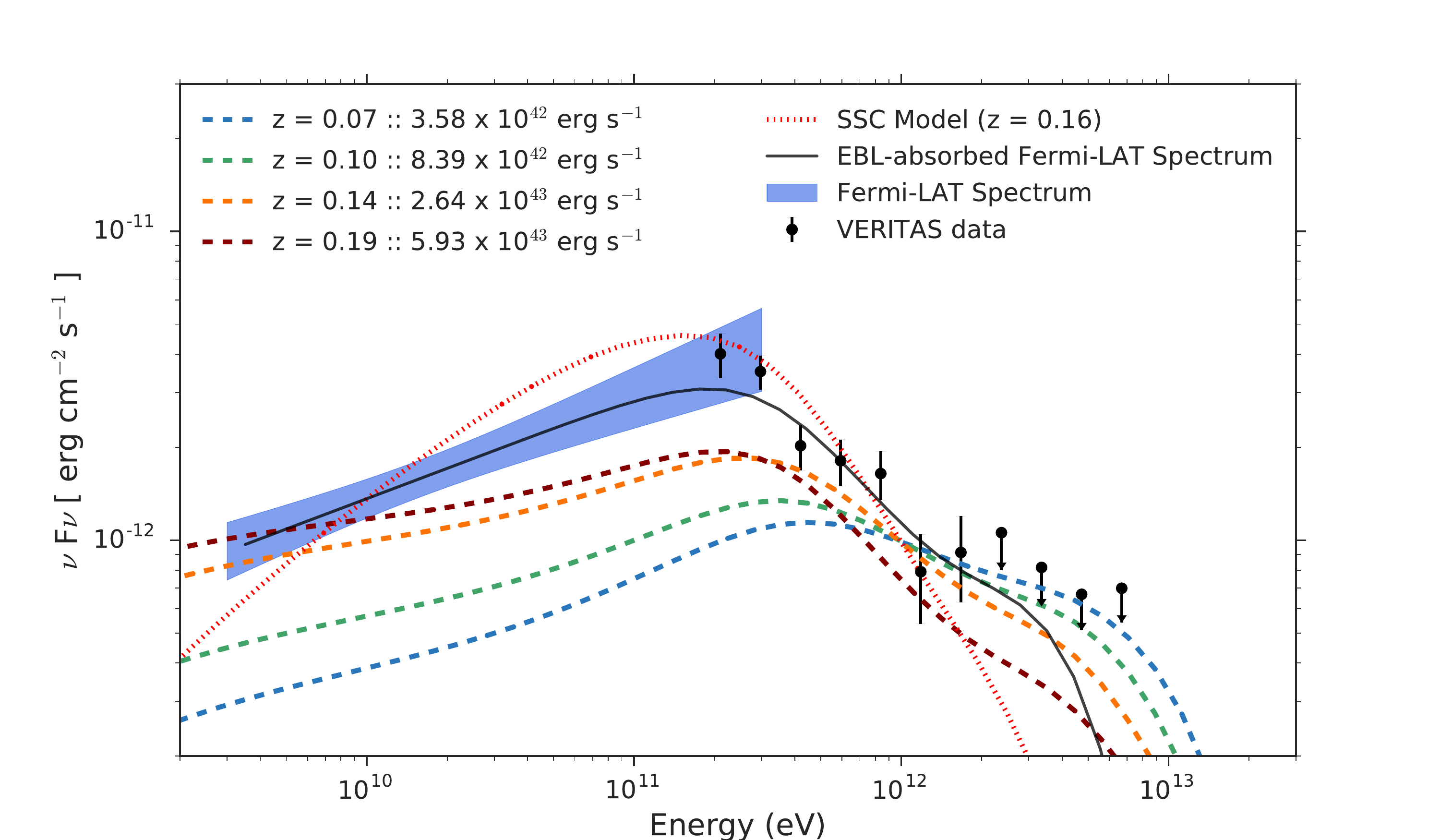}
\caption{Predicted secondary $\gamma$-ray spectra from cascades initiated by UHECRs shown in dashed lines. The legend specifies the assumed distance and the UHECR power required to produce each spectrum. The solid black line shows the \textit{Fermi}-LAT spectrum extrapolated to VHE energies and absorbed by the EBL using the model from \citet{franceschini08}, assuming a source redshift of z = 0.16 (near the best-fit redshift value from the redshift estimation procedure in Section~\ref{sec:redshift}). The dotted red line is the SSC model from Section~\ref{sec:sed}.}
\label{fig:crpropa}
\end{figure}

Due to the uncertain distance of the source, multiple redshifts within the range defined by the lower and upper redshift limits are explored. The two parameters that largely determine the shape of the secondary $\gamma$-ray spectrum are the redshift of the source and the shape of the EBL spectrum. To represent the EBL, the model by \citet{franceschini08} is employed. The list of parameter values used for the simulation is provided in Table~\ref{tab:crpropa}. The choice of intergalactic magnetic field strength, proton spectrum index, and maximum proton energy do not significantly affect the shape of the predicted secondary $\gamma$-ray spectrum, but can change the total cosmic ray power required to produce the secondary $\gamma$-rays by an order of magnitude. Following~\citet{essey10} a Lorentz factor of 10 is assumed, corresponding to a cosmic ray jet opening angle of 6$^{\circ}$. If, instead, the cosmic ray emission is assumed to be isotropic, the power in cosmic rays required to generate the same flux of secondary $\gamma$-rays increases by a factor of 365. 

Figure~\ref{fig:crpropa} illustrates the predicted secondary $\gamma$-ray spectra fit to the VERITAS data, with the condition that the VERITAS and \textit{Fermi}-LAT data are not exceeded by the predicted secondary $\gamma$-ray emission. The resulting requirements on the cosmic ray power for producing the secondary spectra are modest compared to the energy budgets of blazars (the Eddington luminosity of a typical 10$^9$ M$_{\odot}$ SMBH powering a blazar is 10$^{47}$ erg s$^{-1}$). The shapes of the predicted secondary $\gamma$-ray spectra, however, imply that only the highest-energy end of the VERITAS spectrum -- energies $>$ 1 TeV -- can be represented with this type of emission mechanism and only for source distances closer than z $\sim$ 0.14. 

\section{Summary and Conclusions}
\label{sec:conclusions}

HESS J1943+213 is a new addition to the rare class of EHBLs with a strong detection in both HE and VHE $\gamma$-rays. The uncertainty over the source classification has been largely settled in favor of an EHBL with support from results presented here. The detection of a jet-like structure in VLBA 1.6 GHz, 4.3 GHz, and 7.6 GHz bands and the measurements of spectral indices comparable to other known blazars for both the radio core and the jet are strong evidence in support of this position. In addition, the lack of detectable proper motion between EVN and VLBA observations and the two epochs of VLBA observations set constraining upper limits on the transverse velocity as low as 8 km/s if the source is of Galactic origin, essentially ruling out this possibility.

Deep observations with VERITAS detect a source at the HESS J1943+213 position with $\sim$20 $\sigma$ significance, producing a high-statistics spectrum. The VHE spectral properties are consistent with the measurement from H.E.S.S.; however, the VERITAS spectrum extends down to 180~GeV energies, providing an overlap with the \textit{Fermi}-LAT spectrum from eight years of observations. The VERITAS and the \textit{Fermi}-LAT spectra together give an accurate description of the $\gamma$-ray peak of the source SED. These spectra are used to derive more stringent upper limits on the source redshift of z $<$ 0.23.

No statistically significant evidence of flux or spectral variability is found in data from long-term VERITAS observations, as well as in \textit{Swift}-XRT observations over the course of 4 days. As EHBLs are not known for strong variability, the stability of the source is not surprising, but still unusual.

The improved $\gamma$-ray data are used to update and model the broadband SED of HESS J1943+213. An SSC model with a component for the infrared-to-optical light from the host galaxy describes the SED very well, while keeping model parameters to standard values for HBLs. The VLBA data can also be accommodated in the model with the addition of a stratified, conical jet component. Since EHBLs are candidates for hadronic emission, a possible contribution to the $\gamma$-ray portion of the SED from secondary photons produced along the line-of-sight by UHECR-induced cascades is explored for a range of allowed distances for the source. The shape of the secondary $\gamma$-ray spectra, however, makes them viable only for $>$1~TeV energies and only if the source is located closer than z $\sim$ 0.14.

There is still much to learn from HESS J1943+213. High-sensitivity observations of HESS J1943+213 in the hard X-ray band with an instrument like NuSTAR would be valuable for characterizing both the spectral shape and the variability of the emission produced by the higher-energy particles and would help pinpoint the emission mechanism of the source. Moreover, a precise measurement of the distance to HESS J1943+213 would be of great benefit for pinning down its physical properties. With a known distance, the stability of the source combined with its spectral properties in X-rays and $\gamma$-rays would make it an ideal target for studies aiming to constrain the strength of intergalactic magnetic fields and to measure the density of the EBL.

\acknowledgments

This research is supported by grants from the U.S. Department of Energy Office of Science, the U.S. National Science Foundation and the Smithsonian Institution, and by NSERC in Canada. We acknowledge the excellent work of the technical support staff at the Fred Lawrence Whipple Observatory and at the collaborating institutions in the construction and operation of the instrument.

The VERITAS Collaboration is grateful to Trevor Weekes for his seminal contributions and leadership in the field of VHE $\gamma$-ray astrophysics, which made this study possible.

The VLBA is operated by the Long Baseline Observatory. The Long Baseline Observatory is a facility of the National Science Foundation operated under cooperative agreement by Associated Universities, Inc.



\facilities{VERITAS, \textit{Fermi}-LAT, \textit{Swift}-XRT, FLWO 48$''$ (1.2~m), VLBA}

\software{Fermi Science Tools, Fermipy, HEASoft, XSpec, AIPS, CRPropa3}

\bibliography{HESSJ1943_paper}

\end{document}